\documentclass[a4paper,11pt]{article}

\usepackage{jheppub}

\title{\boldmath Superstrings near the conformal boundary of $\rm AdS_3$}

\abstract{We study worldsheet sphere amplitudes, in the RNS formalism, in superstring theory on $\text{AdS}_3\times X$ with pure NS-NS flux using the near-boundary approximation. By computing a number of amplitudes at low-lying spectral flow, we deduce a candidate for a dual CFT for generic $X$. Specialising to $X={\rm S^3}\times \mathbb{T}^4$ and with minimal NS-NS flux, i.e., to the tensionless limit, we explore the effect of the interaction term which we ignore in the near-boundary consideration. We show that amplitudes in the tensionless superstring theory on $\rm AdS_3\times S^3\times\mathbb{T}^4$ do not receive perturbative contributions from such an interaction term.}

\author{Vit Sriprachyakul}
\affiliation{Institut f\"ur Theoretische Physik,
ETH Z\"urich,\\
Wolfgang-Pauli-Strasse 27,
8093 Z\"urich, Switzerland}
\emailAdd{vsriprachyak@phys.ethz.ch}

\usepackage{bm} 
\usepackage{xcolor}
\definecolor{green_maf}{RGB}{28, 166, 46}
\definecolor{blue_mrg}{RGB}{12, 143, 145}
\definecolor{detail}{RGB}{110,110,110}
\usepackage{amsmath} 
\usepackage{nccmath} 
\usepackage{amssymb} 
\usepackage{amsthm} 
\usepackage{mathtools} 
\usepackage[utf8]{inputenc} 
\usepackage{braket}
\usepackage{enumerate}
\usepackage[inline]{enumitem}
\usepackage{comment}
\usepackage{soul} 
\usepackage{tikz}
\usepackage{csquotes}
\newcommand{\RR}{\mathbb{R}}

\usetikzlibrary{calc,decorations.markings}
\usetikzlibrary{decorations.pathmorphing}
\usetikzlibrary{shapes,backgrounds}
\usetikzlibrary{fadings}
\usetikzlibrary{cd}
\usetikzlibrary{decorations.pathreplacing,calligraphy}

\tikzset{
partial ellipse/.style args={#1:#2:#3}{
insert path={+ (#1:#3) arc (#1:#2:#3)}
}
}

\newif\ifdetails
\detailstrue

\def\be{\begin{equation}}
\def\ee{\end{equation}}

\begin{document}

\maketitle
\newpage

\section{Introduction}

$\rm AdS_3/CFT_2$ has been a great playground for understanding the $\rm AdS/CFT$ duality \cite{Maldacena:1997re,Witten:1998qj}. This is due to the fact that 2-dimensional conformal field theories (CFTs) have a tremendously larger number of symmetries than their higher dimensional cousins. Moreover, the worldsheet theory with pure NS-NS flux splits into purely left/right degrees of freedom and the worldsheet sigma model is described by a WZW model. Despite this simplification, deriving CFT correlators directly from the string worldsheet theory on $\rm AdS_3$ is hard. However, huge progress has recently been made in computing string correlators in $\rm AdS_3$ with pure NS-NS flux. In particular, $2$-point, $3$-point and $4$-point functions have been calculated \cite{Dei:2021xgh,Dei:2021yom,Dei:2022pkr,Bufalini:2022toj,Iguri:2022pbp,Fiset:2022erp,Eberhardt:2021vsx} (see also \cite{Maldacena:2001km,Gaberdiel:2007vu,Giribet:2008yt,Cardona:2009hk,Cardona:2010qf,Halder:2022ykw,Teschner:1997ft,Teschner:1999ug,Giribet:1999ft,Giribet:2000fy,Giribet:2001ft,Cagnacci:2013ufa,Hikida:2020kil,Hikida:2023jyc} for related works) and it was shown that residues in string amplitudes match the perturbative expansion in the dual CFT \cite{Eberhardt:2021vsx,Dei:2022pkr} to low orders in perturbation theory. More recently in \cite{Knighton:2023mhq,Knighton:2024qxd}, an efficient way to compute these residues was found and perturbative dual CFT correlators were derived directly from the worldsheet. Thus, this adds an extremely valuable understanding to the holographic principle. Nevertheless, the calculations in \cite{Knighton:2024qxd} only concern bosonic strings in $\rm AdS_3$ and therefore, supersymmetric extension is highly desirable.

The computation in \cite{Knighton:2024qxd} was made possible by adopting a certain approximation in the worldsheet theory. The approximation that allows us to extract the residues of the worldsheet amplitudes is dubbed the `near-boundary limit' \cite{Knighton:2024qxd,Knighton:2023mhq,Giveon:2001up,Giveon:1998ns}. Intuitively, this amounts to considering a contribution in which the worldsheet stays near the conformal boundary of $\rm AdS_3$. This cartoon is also known as the `worldsheet instanton' in \cite{Maldacena:2001km}. The magic of this limit is that the sigma model on the worldsheet becomes free \cite{Knighton:2024qxd,Knighton:2023mhq,Giveon:1998ns,Giveon:2001up} and thus worldsheet amplitudes can be evaluated analytically in the path integral formalism. In order to make use of the power of the near-boundary approximation in superstring theory, we perform the well-known trick of decoupling the $\rm AdS_3$ fermions from the $SL(2,\RR)$ currents. We then apply the near-boundary result to compute correlators in the $SL(2,\RR)$ bosonic current theory. There is no real challenge in computing correlators in the fermion theory since this can be done by the usual Wick contractions. However, we will see that writing fermion correlators in terms of holographic (covering map) data is nontrivial and is the main obstacle that prevents us from calculating generic $n$-point functions. Aside from this difficulty, we see that superstring amplitudes can be computed given the result from the near-boundary bosonic string theory. In the main text, we will provide details of how vertex operators in various picture numbers should be constructed and we will show by explicit worldsheet calculations that sensible holographic dictionary can be extracted.

We would like to emphasise that the $\rm AdS_3$ fermions are crucial and are in fact at the centre of our discussion. Let us briefly explain why. From \cite{Knighton:2024qxd}, correlators in the bosonic $\mathfrak{sl}(2,\mathbb{R})_k$ WZW model reproduce correlators in a deformed symmetric orbifold theory, with the seed theory of central charge $6k$. However, the worldsheet theory that describes superstrings in $\rm AdS_3$ is a supersymmetric $\mathcal{N}=1$ $\mathfrak{sl}^{(1)}(2,\mathbb{R})_k$ WZW model that consists of the currents $J^3,J^\pm$ and their superpartners $\psi^3,\psi^\pm$. This supersymmetric theory can be written as a sum of two decoupled theories: bosonic $\mathfrak{sl}(2,\mathbb{R})_{k+2}$ theory (with currents $\mathcal{J}^3,\mathcal{J}^\pm$ to be defined later) and three (free) fermions. The residues in the superstring worldsheet amplitudes are obtained from the residues in the $\mathfrak{sl}(2,\mathbb{R})_{k+2}$ theory which we compute using the near-boundary limit. This implies that, if we completely ignore the fermions, the dual CFT is a perturbed symmetric orbifold with a seed theory of central charge $6(k+2)=6k+12$. However, this is not the right central charge since it was known to be $6k$ \cite{Argurio:2000tb}. We will see more in Section \ref{sec:correlator} how the fermions rectify the purely bosonic answer and give rise to consistent superstring holography. Explicitly, a duality we will gain evidences for is 
\begin{equation}
\begin{aligned}
&\text{superstrings in }{\rm AdS_3}\times X\text{ near the boundary}\\
&\hspace{4cm}\Longleftrightarrow\\
&{\rm Sym}^\infty\left(\RR_{\mathcal{Q}}^{(1)}\times X\right)+\text{a marginal twist-}2\text{ deformation}\,,
\label{eq:our proposal}
\end{aligned}
\end{equation}
where $\RR_{\mathcal{Q}}^{(1)}:=\RR_{\mathcal{Q}}\times(\text{a free fermion})$ and $\RR_{\mathcal{Q}}$ is a linear dilaton theory with background charge $\mathcal{Q}$ that will be determined in Section \ref{sec:correlator}. The CFT should be treated perturbatively in a similar sense as in \cite{Eberhardt:2021vsx,Knighton:2024qxd}. We emphsise again that our proposed duality is for any consistent (at least $\mathcal{N}=1$) supersymmetric background $X$.

It is clear that the near-boundary approximation cannot completely describe the worldsheet theory with generic NS-NS flux. However, there is a special value of the flux where we can expect the near-boundary approximation to be exact. This is the so-called tensionless ($k=1$) limit of superstring theory. In $\rm AdS_3\times S^3\times \mathbb{T}^4$ string theory with the minimal amount of NS-NS flux, it was shown that the worldsheet theory has free field realisations \cite{Eberhardt:2018ouy,Dei:2023ivl}. Indeed, we will show in our paper that the near-boundary (free field) approximation is perturbatively exact, that is, the perturbative expansion of the interaction term we ignore in going to the near-boundary limit is vanishing to all order except at zeroth order. Indeed, that the near-boundary approximation becomes exact in the tensionless limit is expected and was speculated to be true \cite{McStay:2023thk,Dei:2023ivl}. However, there was no explicit derivation of this claim and we now give one such derivation.\footnote{We should note that there was an argument in the hybrid formulation of superstrings \cite{Dei:2023ivl} which makes use of the fact that the interaction term is not in the physical spectrum and thus could somehow be ignored. However, the authors of \cite{Dei:2023ivl} did not discuss the effect of including this term in the action.}

The paper is organised as follows. In Section \ref{sec:rnsstring}, we review the RNS formalism of superstrings propagating in ${\rm AdS_3}\times X$. We briefly discuss how correlators are calculated. In Section \ref{sec:near boundary review}, we review the near-boundary approximation in the purely bosonic case as well as adapting it to the supersymmetric setup. In Section \ref{sec:correlator}, we compute a number of $3$-point and $4$-point amplitudes and discuss the implications on the dual CFT. We discuss in Section \ref{sec:tensionless limit} the effect the interaction term has in the tensionless limit. 
We end our exposition in Section \ref{sec:discussion} with discussions and interesting future directions. Appendices \ref{sec:all tedious calculations} and \ref{sec:betagamma with interaction} collect some technical results.

\section{\boldmath Superstrings in \texorpdfstring{$\text{AdS}_3\times X$}{AdS3 times X} in the RNS formalism}\label{sec:rnsstring}

Throughout this paper we will use the RNS formalism when discussing worldsheet correlators. We will follow quite closely the general discussion in \cite{Polchinski:1998rr}, while a specific discussion of the RNS formalism applied to strings in $\rm AdS_3$ will mostly follow \cite{Ferreira:2017pgt}. In this paper, we will focus on the left-moving degrees of freedom for simplicity. 

\subsection[\texorpdfstring{$\rm AdS_3$}{AdS3} worldsheet theory]{\boldmath \texorpdfstring{$\rm AdS_3$}{AdS3} worldsheet theory}
The RNS treatment starts from considering a tensor product $\mathfrak{sl}(2,\RR)^{(1)}_k\otimes X$. This is the worldsheet theory that describes superstrings propagating in the background ${\rm AdS_3}\times X$. The superscript indicates that we have an $\mathcal{N} =1$ supersymmetric $\mathfrak{sl}(2,\RR)$ WZW model.\footnote{We will also assume that the compact CFT $X$ has at least an $\mathcal{N}=1$ supersymmetry.} The fermions in the $\mathfrak{sl}(2,\RR)^{(1)}_k$ theory live in the adjoint representation of the current algebra. The full worldsheet theory consists of the tensor product $\mathfrak{sl}(2,\RR)^{(1)}_k\otimes X$ as the matter part and the usual superconformal ghosts $b,c,\tilde \beta,\tilde \gamma$. It is often useful to decouple the fermions and the currents in $\mathfrak{sl}(2,\RR)^{(1)}_k$. Decoupling the fermions results in two decoupled theories: the bosonic $\mathfrak{sl}(2,\RR)_{k+2}$ WZW model and three free fermions. Let us now look at the decoupling more closely.

The generators in $\mathfrak{sl}(2,\RR)^{(1)}_k$ are the bosonic currents $J^3,J^\pm$ and the (adjoint) fermions $\psi^3,\psi^\pm$. They satisfy the following commutators
\begin{equation}
\begin{aligned}
[J^+_m,J^-_n]=&-2J^3_{m+n}+km\delta_{m+n,0},& [J^3_m,J^\pm_n]=&\pm J^\pm_{m+n},& [J^3_m,J^3_n]=&-\frac{k}{2}m\delta_{m+n,0},\\
[J^\pm_m,\psi^3_r]=&\mp \psi^\pm_{m+r},& [J^3_m,\psi^\pm_r]=&\pm \psi^\pm_{m+r},& [J^\pm_m,\psi^\mp_r]=&\mp2\psi^3_{m+r},\\
\{\psi^+_r,\psi^-_s\}=&k\delta_{r+s,0}& \{\psi^3_r,\psi^3_s\}=&-\frac{k}{2}\delta_{r+s,0}\,.&
\label{commu}
\end{aligned}
\end{equation}
To decouple the fermions we consider new currents defined by
\begin{equation}
\begin{aligned}
\mathcal{J}^+:=&J^++\frac{2}{k}\psi^3\psi^+\,,\\
\mathcal{J}^-:=&J^--\frac{2}{k}\psi^3\psi^-\,,\\
\mathcal{J}^3:=&J^3+\frac{1}{k}\psi^-\psi^+\,.
\end{aligned}
\end{equation}
One can check directly that $\mathcal{J}^a$ and $\psi^a$ have trivial commutators using \eqref{commu}. Note that the decoupled currents $\mathcal{J}^a$ form a Kac-Moody algebra of $\mathfrak{sl}(2,\RR)$ with level $k+2$ as can be seen, for example, from the commutator $[\mathcal{J}^3_m,\mathcal{J}^3_n]$. Since the $\mathfrak{sl}(2,\RR)^{(1)}_k$ theory posseses an $\mathcal{N}=1$ supersymmetry, there exists a supercurrent and it is given by
\begin{equation}
G^{sl}=\frac{1}{k}\left( 
\mathcal{J}^+\psi^-+\mathcal{J}^-\psi^+-2\mathcal{J}^3\psi^3-\frac{2}{k}\psi^+\psi^-\psi^3 \right)\,.
\label{eq:ade3G}
\end{equation} 
The full matter supercurrent is given as a sum of \eqref{eq:ade3G} and the supercurrent $G^X$ from the $X$ theory. The stress energy tensor of the $\rm AdS_3$ part of the theory can be constructed as a sum of the stress tensor in the $\mathfrak{sl}(2,\RR)_{k+2}$ theory and the stress tensor of the fermion theory. However, we shall not need its explicit form in our discussion. As usual in the theory that contains fermions, it consists of two sectors depending on the integrality of the mode number of fermions. The half-integer moded sector is the NS sector and the integer moded sector is the R sector. The ground states in each sector will be defined and we analyse them in turn.

\subsubsection*{NS sector}
In the NS sector, the ground state of the fermion theory is just the vacuum, annihilated by all fermionic positive modes, whereas the ground states in the decoupled current $\mathcal{J}^a$ theory are constructed from the irreducible representations of $\mathfrak{sl}(2,\RR)$. There are two parameters, $m$ and $j$, parametrising the ground states of the current theory and depending on the range of the parameters, the representations they furnish are either the discrete representations $\mathcal{D}^{\pm}_j$ or the continuous representations $\mathcal{C}_{\alpha}^j$. Since we will not discuss the string physical spectrum, we shall not go into detail of how these representations are defined. Interested readers may consult \cite{Maldacena:2000hw,Knighton:2024qxd,Eberhardt:2018ouy} for more details. Here, $m$ denotes the eigenvalue of $\mathcal{J}^3_0$ and $j$ is the $\mathfrak{sl}(2,\RR)$ spin. Explicitly, we have
\begin{equation}
\begin{aligned}
\mathcal{J}^3_0\ket{m,j}=m\ket{m,j}\,,\quad \mathcal{J}^{\pm}_{0}\ket{m,j}=(m\pm j)\ket{m\pm 1,j}\,.
\end{aligned}
\end{equation}
Thus, the ground states in the NS sector is of the form 
\begin{equation}
\ket{m,j}^{NS}:=\ket{m,j}\otimes\ket{0}\,,
\end{equation}
where $\ket{0}$ is the vacuum state in the fermion theory. The (unflowed) Hilbert space is spanned by the states above and also by the $\mathcal{J}^a_{-n},\psi^a_{-r}$ descendants of the ground states.

\subsubsection*{R sector}
In the Ramond sector, the ground states in the $\rm AdS_3$ fermion theory form an irreducible representation of the Clifford algebra in $2+1$ dimensions which is two dimensional. Explicitly, we adopt the following convention
\begin{equation}
\begin{aligned}
\psi^+_0\ket{0}_{R,\pm}=0\,,
\label{eq:R sector ground states conditions}
\end{aligned}
\end{equation}
where the subscripts denote the theory the ground states belong to. That is $\ket{0}_{R,\pm}$ lives in the $\psi^\pm$ theory. The R ground states of the remaining fermions, which are $\psi^3$ and the fermions from $X$, are denoted collectively by $\ket{0}_{R,\text{rest}}$. This split will make the discussion on the holography more transparent in Section \ref{sec:correlator}. For readability, we define a notation
\begin{equation}
\ket{m,j}^R:=\ket{m,j}\otimes\ket{0}_{R,\pm}\,.
\end{equation}
Note also that because of \eqref{eq:R sector ground states conditions}, the eigenvalues under $J^3_0$ and $\mathcal{J}^3_0$ are different. The eigenvalue of $\ket{m,j}^R$ under $\mathcal{J}^3_0$ is still $m$ but
\begin{equation}
\begin{aligned}
J^3_0\ket{m,j}^R=&\left( \mathcal{J}^3_0-\frac{1}{2k}(\psi^-_0\psi^+_0-\psi^+_0\psi^-_0) \right)\ket{m,j}^R\\
=&\left(m+\frac{1}{2}\right)\ket{m,j}^R\,,\\
J^3_0\psi^-_0\ket{m,j}^R=&\left( \mathcal{J}^3_0-\frac{1}{2k}(\psi^-_0\psi^+_0-\psi^+_0\psi^-_0) \right)\psi^-_0\ket{m,j}^R\\
=&\left(m-\frac{1}{2}\right)\psi^-_0\ket{m,j}^R\,.
\end{aligned}
\end{equation}
This shift will play an important role when we discuss the holographic dictionary.

\subsubsection{Spectrally flowed states}
As explained in \cite{Maldacena:2000hw,Maldacena:2000kv,Maldacena:2001km}, in order to capture strings winding around the boundary of $\rm AdS_3$, we need to include in the full string Hilbert space states that are obtained from the spectral flow action. These are the so-called \emph{spectrally flowed states}. To define the action of spectral flow on the states, we first define its action on the currents and the fermions. Note that the action of spectral flow is an automorphism of the Kac-Moody algebra and thus preserves the form of the commutators \eqref{commu}. We define the spectral flow to act as follows
\begin{equation}
\begin{aligned}
\sigma^w(\mathcal{J}^\pm_n)=&\mathcal{J}^\pm_{n\mp w}\,,&\quad \sigma^w(\mathcal{J}^3_n)=&\mathcal{J}^3_{n}+\frac{(k+2)w}{2}\delta_{n,0}\,,\\
\sigma^w(\psi^\pm_n)=&\psi^\pm_{n\mp w}\,,&\quad
\sigma^w(\psi^3_n)=&\psi^3_{n}\,,
\label{eq:spectral flow action on fields}
\end{aligned}
\end{equation}
where $w$ denotes the unit of spectral flow with $w=0$ meaning no spectral flow.\footnote{At this point, one may choose not to flow the fermions above since the two theories decouple, the motivation behind our choice has to do with the fact that \eqref{eq:spectral flow action on fields} preserves the supercurrent up to $\psi^3$ that is
\begin{equation}\nonumber
\sigma^w(G^{sl}_r)=G^{sl}_r+2w\psi^3_r\,.
\end{equation}
Intuitively, this ensures that the $\mathcal{N}=1$ supersymmetry on the worldsheet is respected.}

The states in the left-moving Hilbert space (before physical projection) therefore consist of the ground states, the states obtained by acting on the ground states by raising operators, i.e., negative modes of $\mathcal{J}^a$ and $\psi^a$ and the spectral flow images of those states. By combining the left-moving flowed module with the right-moving flowed module in a modular invariant way, one can form a consistent string spectrum on the worldsheet. We will not go into details of the spectral analysis nor discuss the physical Hilbert space in depth. It has been explored for the case of superstrings in $\rm AdS_3\times S^3\times\mathbb{T}^4$ in \cite{Ferreira:2017pgt,Eberhardt:2019qcl}, for bosonic strings in ${\rm AdS_3}\times X$ in \cite{Eberhardt:2019qcl}, for superstrings at $k=1$ in ${\rm AdS_3}\times X$ in \cite{Giribet:2018ada,Gaberdiel:2018rqv} and for superstrings at generic $k$ in ${\rm AdS_3}\times X$ in \cite{Argurio:2000tb}. Since every state in the string theory are constructed from the action of raising operators and spectral flow on the ground states, the most basic correlators are correlators of $w-$spectrally flowed ground states whose properties we now discuss. 

A $w-$spectrally flowed image of a state $\ket{\Psi}$ is denoted by $[\ket{\Psi}]^{\sigma^w}$ and is defined to satisfy the following condition
\begin{equation}
A_m[\ket{\Psi}]^{\sigma^w}=\left[\sigma^w(A_m)\ket{\Psi}\right]^{\sigma^w}\,,
\label{defspecflow}
\end{equation}
where $A_m$ collectively denotes the modes of $\mathcal{J}^a$ and $\psi^a$. For the NS sector, a $w-$spectrally flowed image of $\ket{m,j}^{NS}$ is denoted by $\left[\ket{m,j}^{NS}\right]^{\sigma^w}$. Using the definition \eqref{defspecflow}, this state satisfies
\begin{equation}
\begin{aligned}
\mathcal{J}^3_0\left[\ket{m,j}^{NS}\right]^{\sigma^w}&=\left(m+\frac{(k+2)w}{2}\right)\left[\ket{m,j}^{NS}\right]^{\sigma^w}\,,\\
\mathcal{J}^3_n\left[\ket{m,j}^{NS}\right]^{\sigma^w}&=0\,,\quad n>0,\\
\mathcal{J}^{\pm}_{n}\left[\ket{m,j}^{NS}\right]^{\sigma^w}&=0\,,\quad n>\pm w\,,\\
\psi^\pm_r\left[\ket{m,j}^{NS}\right]^{\sigma^w}&=0\,,\quad r>\pm w,\\
\psi^3_r\left[\ket{m,j}^{NS}\right]^{\sigma^w}&=0\,,\quad r>0\,.
\label{eq:spectral flow condition NS sector}
\end{aligned}
\end{equation}
On the other hand, for the R sector, a $w-$spectrally flowed image of $\ket{m,j}^{R}$ is denoted by $\left[\ket{m,j}^{R}\right]^{\sigma^w}$ and an analogue of \eqref{eq:spectral flow condition NS sector} can be found by using \eqref{eq:spectral flow action on fields} and \eqref{eq:R sector ground states conditions} to be
\begin{equation}
\begin{aligned}
\psi^+_r\left[\ket{m,j}^{R}\right]^{\sigma^w}&=0\,,\quad r\geq w,\\
\psi^-_r\left[\ket{m,j}^{R}\right]^{\sigma^w}&=0\,,\quad r>- w,\\
\psi^3_r\ket{0}_{R,\text{rest}}&=0\,,\quad r>0\,.
\label{eq:spectral flow condition R sector}
\end{aligned}
\end{equation}
Since the current ground states are the same in $\ket{m,j}^{NS}$ and $\ket{m,j}^{R}$, the action of the currents $\mathcal{J}^a_n$ is the same as in \eqref{eq:spectral flow condition NS sector}.

\subsubsection*{Physical state condition}

Although we will not discuss in detail how to compute the string physical spectrum, we give a short summary of the procedure and its consequences. In the old covariant quantisation, one first starts by imposing the physical state condition,  that is, that the positive modes of $L^{\text{matter}}$ and $G^{\text{matter}}$ annihilate the physical states. Then, the negative modes generate null states which have to be quotiented out. The condition from the zero modes finally give rise to the mass shell condition. In the modern covariant quantisation, the superconformal ghosts are introduced and the physical states are obtained by performing a BRST approach. At the end of this process, two bosonic and two fermionic degrees of freedom are removed. In the bosonic string, only the condition on $L^{\text{matter}}$ is imposed and thus two bosonic degrees of freedom are eliminated. This is consistent with the analysis of \cite{Knighton:2024qxd} that the worldsheet amplitudes give rise to a dual CFT that consists of two less bosonic degrees of freedom than that of the worldsheet theory. In superstrings, a similar phenomenon happens, that is, the dual CFT has two bosonic and two fermionic degrees of freedom less than that of the worldsheet theory. Since both sides of the duality involve the same compact theory $X$, we focus on the remaining part. Indeed, the worldsheet theory has 3 bosonic and 3 fermionic fields from the $\rm AdS_3$ part whereas the dual CFT \eqref{eq:our proposal} only has 1 bosonic and 1 fermionic fields from $\RR^{(1)}_{\mathcal{Q}}$.

\subsection{RNS precription of correlators}

Let $V(z)$ be a superconformal primary matter field, that is, $V(z)$ is annihilated by the positive modes of the matter stress tensor and the matter supercurrent
\begin{equation}
L_m\cdot V=0,\quad G_r\cdot V=0,\quad  m,r>0\,.
\end{equation}
Then, if $V$ is in the NS sector, the $0-$picture vertex operator is given by 
\begin{equation}
V^0=G_{-\frac{1}{2}}\cdot V\,,
\end{equation}
and the $(-1)-$picture NS sector vertex operator is given by 
\begin{equation}
V^{-1}=e^{-\phi}V\,.
\end{equation}
Note that to go from the $(-1)$-picture vertex operator to its $0$-picture form, we simply have to apply the picture-changing-operator (PCO), see also \cite{Polchinski:1998rr,Friedan:1985ge} for more details. If instead $V$ is in the R sector, the natural object to consider is the $\left(-\tfrac{1}{2}\right)-$picture vertex operator which is given by
\begin{equation}
V^{-\frac{1}{2}}=e^{-\frac{\phi}{2}}V\,.
\end{equation}
Here, $\phi$ comes from bosonising the superconformal $\tilde\beta\tilde\gamma$ ghosts as 
\begin{equation}
\tilde\beta=e^{-\phi}\partial\xi, \quad\tilde\gamma=e^\phi\eta
\end{equation}
with $\phi(z)\phi(w)\sim-\ln(z-w)$ and $\xi,\eta$ fermionic ghosts satisfying 
\begin{equation}
\eta(z)\xi(w)\sim\frac{1}{z-w}\,.
\end{equation}
A genus zero $n$-point correlator is given by the expression
\begin{equation}
\Braket{\left( c(z_1)V^{-1}(z_1) \right)\left( c(z_2)V^{-1}(z_2) \right)\left( c(z_3)V^{0}(z_3) \right)\left( V^{0}(z_4) ... V^0(z_n) \right)}\,,
\end{equation}
if all the insertions are in the NS sector. For our purposes, correlators with two insertions from the R sector are given by
\begin{equation}
\Braket{\left( c(z_1)V^{-\frac{1}{2}}(z_1) \right)\left( c(z_2)V^{-\frac{1}{2}}(z_2) \right)\left( c(z_3)V^{-1}(z_3) \right)\left( V^{0}(z_4) ... V^0(z_n) \right)}\,.
\end{equation}
A worldsheet amplitude is obtained by integrating these expressions over $z_4,...,z_n$. We will later show that this prescription together with the near-boundary approximation gives rise to amplitudes that can be interpreted naturally as deformed symmetric orbifold correlators. Note that this distribution of picture number is entirely for later convenience. As long as one saturates the picture number conservation, one may freely move around the picture-changing-operators \cite{Friedan:1985ge,Polchinski:1998rr}.

\section{\boldmath Near-boundary approximation of strings in \texorpdfstring{$\rm AdS_3$}{AdS3}}\label{sec:near boundary review}
We saw in the previous section that the supersymmetric $\mathfrak{sl}(2,\RR)^{(1)}_k$ WZW model can be decomposed into two decoupled theories. Thus the building blocks of correlation functions in the supersymmetric theory are correlators in the bosonic current theory and the correlators in the free fermion theory. The correlators in the free fermion theory can be computed straightforwardly by Wick contractions but the correlators in the current theory are more complicated \cite{Dei:2022pkr,Dei:2021yom,Dei:2021xgh}. Nonetheless, as far as holography is concerned, it was shown that the holographically important information can be extracted from the residues\footnote{More precisely, it is the residues of the \emph{bulk poles} \cite{Eberhardt:2021vsx} that can be identified with the perturbative structure of the dual CFT.} in the current correlators \cite{Eberhardt:2021vsx,Dei:2022pkr}. Furthermore, it was shown in \cite{Knighton:2024qxd} that these residues can be computed explicitly in the near-boundary approximation without having to calculate the full $SL(2,\RR)$ correlation functions. These residues are simply the near-boundary amplitudes without the (radial) momentum conservation delta function. In this section, we review the near-boundary approximation while at the same time discussing the supersymmetric version of it. Our bosonic discussion is based mainly on \cite{Knighton:2023mhq,Knighton:2024qxd}.

\subsection[Near-boundary approximation of strings in \texorpdfstring{$\rm AdS_3$}{AdS3}]{\boldmath Near-boundary approximation of strings in \texorpdfstring{$\rm AdS_3$}{AdS3}}
Near-boundary approximation of the $\mathfrak{sl}(2,\RR)_{k+2}$ current theory starts from considering the first order action
\begin{equation}
S_{\rm AdS_3}=\frac{1}{2\pi}\int d^2z\left( \frac{1}{2}\partial\Phi\bar\partial\Phi-\frac{QR\Phi}{4}+\beta\bar\partial\gamma+\bar\beta\partial\bar\gamma-\nu\beta\bar\beta e^{-Q\Phi} \right)\,,
\label{eq:ads full action}
\end{equation}
where $Q=\sqrt{2/(k_b-2)}$, $k_b:=k+2$ and $\nu$ some coupling constant. We then ignore the interaction term $\beta\bar\beta e^{-Q\Phi}$ in order to go near the boundary. This is because the boundary of $\rm AdS_3$ is at $\Phi=+\infty$. The $\mathfrak{sl}(2,\RR)_{k_b}$ currents can be written in terms of the free fields as follows
\begin{equation}
\begin{aligned}
\mathcal{J}^+=\beta\,,\quad \mathcal{J}^3=-\frac{1}{Q}\partial\Phi+\beta\gamma\,,\quad \mathcal{J}^-=-\frac{2}{Q}\partial\Phi\,\gamma+\beta\gamma^2-k_b\partial\gamma\,,
\label{eq:defWaki}
\end{aligned}
\end{equation}
where normal-ordering is implicitly understood. This construction is the so-called Wakimoto representation. As we mentioned earlier, we will focus on the chiral (holomorphic) part of the theory, however, it should be clear from our construction that the right-moving objects can be obtained similarly. The stress tensor in terms of free fields is given by
\begin{equation}
T=-\frac{1}{2}\partial\Phi\partial\Phi-\frac{Q}{2}\partial^2\Phi-\beta\partial\gamma\,,
\end{equation}
and the central charge is simply a sum of the central charges of $\Phi$ and $\beta\gamma$ theories which is
\begin{equation}
c({\rm AdS_3})=1+3Q^2+2=\frac{3k_b}{k_b-2}=\frac{3(k+2)}{k}\,.
\end{equation}
The OPEs between the free fields are given as usual as
\begin{equation}
\Phi(z)\Phi(w)\sim -\ln|z-w|^2,\quad \beta(z)\gamma(w)\sim-\frac{1}{z-w}\,.
\end{equation}
One can check that this does give rise to the correct $\mathfrak{sl}(2,\RR)_{k+2}$ algebra. The near-boundary approximation does not affect the fermions. In particular, they remain free and decoupled from the $\Phi,\beta,\gamma$ fields and their OPEs are unchanged. 

\subsection{Spectrally flowed vertex operators}
Having a free field realisation of the currents enables us to write down a useful form of the spectrally flowed ground state vertex operators. This is given, for the current theory, by \cite{Knighton:2023mhq,Knighton:2024qxd}
\begin{equation}
V\left( [\ket{m,j}]^{\sigma^w} ,z\right):=V^w_{m,j}(z)=e^{(w/Q-Qj)\Phi}\bigg(\frac{\partial^w\gamma}{w!}\bigg)^{-m-j}\delta_w(\gamma)\,,
\label{eq:def bosonic vertex operator}
\end{equation}
where $\delta_w(\gamma)$ is defined as follows
\begin{equation}
\delta_w(\gamma):=\prod_{i=0}^{w-1}\delta\left(\frac{\partial^i\gamma}{i!}\right)\,.
\label{eq:definition of delta function in bosonic theory}
\end{equation}
The fermion part of the vertex operator can also be found by noting that the condition \eqref{eq:spectral flow condition NS sector} in the NS sector translates to
\begin{equation}
\begin{aligned}
\psi^\pm(z)V\left( [\ket{0}]^{\sigma^w} ,0\right)=z^{\mp w}V\left( \left[\psi^\pm_{-\frac{1}{2}}\ket{0}\right]^{\sigma^w} ,0\right)+\cdots\,,
\end{aligned}
\end{equation}
where $\cdots$ denotes higher order terms. By bosonising the $\psi^\pm$ fermions as
\begin{equation}
\psi^\pm=\sqrt{k}e^{\pm iH}\,,
\end{equation}
with $H(z)H(w)\sim -\ln(z-w)$, the spectrally flowed vertex operator can be written as 
\begin{equation}
V\left( [\ket{0}]^{\sigma^w}\right)=e^{-iwH}\,.
\end{equation}
Refermionising this expression gives, for $w>0$,\footnote{Strictly speaking, we should write $e^{-iwH}=k^{-\frac{w}{2}}\prod_{i=0}^{w-1}\frac{\partial^i\psi^-}{i!}$ with the prefactor to account for the fact that $\psi^\pm=\sqrt{k}e^{\pm iH}$. However, two vertex operators differ by an overall constant define the same state and we can thus \emph{define} $V\left( [\ket{0}]^{\sigma^w}\right):=\prod_{i=0}^{w-1}\frac{\partial^i\psi^-}{i!}$ and we consistently use this definition in our subsequent computations.}
\begin{equation}
e^{-iwH}=\prod_{i=0}^{w-1}\frac{\partial^i\psi^-}{i!}=\prod_{i=0}^{w-1}\delta\left(\frac{\partial^i\psi^-}{i!}\right)=:\delta_w\left(\psi^-\right)\,.
\label{eq:refermionising}
\end{equation}
We note that despite the similarity between the above expression and \eqref{eq:definition of delta function in bosonic theory}, the spectral flow action has vastly different effects on the bosonic current theory and the fermionic theory. In the bosonic current theory, spectral flow generates a new, non-highest weight representations with respect to the current algebra. While in the fermionic theory, it simply maps a state in the theory to another and does not generate any new representations. In particular, the state \eqref{eq:refermionising} is just a descendant in the fermion theory.

The R sector spectrally flowed vertex operator can also be found by similar calculations and the results are
\begin{equation}
\begin{aligned}
V\left( [\ket{0}_{R,\pm}]^{\sigma^w},z\right)=&e^{-i\left(w-\frac{1}{2}\right)H(z)}\,,\\
V\left( [\psi^-_0\ket{0}_{R,\pm}]^{\sigma^w},z\right)=&e^{-i\left(w+\frac{1}{2}\right)H(z)}\,.
\label{eq:spectral flow in R sector with no x translation}
\end{aligned}
\end{equation}
We also see that by including the spectrally flowed sector, the states in the first line and the states in the second line of \eqref{eq:spectral flow in R sector with no x translation} overlap. One may worry that this means we cannot distinguish the overlapping states of the first two lines inside a correlator. Indeed, we will see that they are indistinguishable unless we evaluate the correlator on the covering map. 

Putting the current and fermion parts together, we obtain the spectrally flowed ground state vertex operator
\begin{equation}
\begin{aligned}
V^{w,NS}_{m,j}=:V\left( \left[\ket{m,j}^{NS}\right]^{\sigma^w} ,z\right)=&e^{(w/Q-Qj)\Phi}\bigg(\frac{\partial^w\gamma}{w!}\bigg)^{-m-j}\delta_w(\gamma)e^{-iwH}\\
=&e^{(w/Q-Qj)\Phi}\bigg(\frac{\partial^w\gamma}{w!}\bigg)^{-m-j}\delta_w(\gamma)\delta_w(\psi^-)\\
=&V^w_{m,j}\delta_w(\psi^-)\,,
\label{eq:general vertex operator NS sector}
\end{aligned}
\end{equation}
for the NS sector vertex operator and 
\begin{equation}
\begin{aligned}
V\left( \left[\ket{m,j}^{R}\right]^{\sigma^w}\otimes\ket{0}_{R,\text{rest}} ,z\right)=&e^{(w/Q-Qj)\Phi}\bigg(\frac{\partial^w\gamma}{w!}\bigg)^{-m-j}\delta_w(\gamma)e^{-i\left(w-\frac{1}{2}\right)H}V\left( \ket{0}_{R,\text{rest}},z\right)\\
=&V^w_{m,j}e^{-i\left(w-\frac{1}{2}\right)H}V\left( \ket{0}_{R,\text{rest}},z\right)\,,
\label{eq:general vertex operator R sector}
\end{aligned}
\end{equation}
for the R sector vertex operator. In the next section we will see that this is an appropriate matter vertex operator in the $(-1)-$picture (for the NS sector) and $\left(-\tfrac{1}{2}\right)-$picture (for the R sector), when paired with $e^{-\phi},e^{-\frac{1}{2}\phi}$, respectively. Since we also need $0-$picture NS sector vertex operators in our computations, we derive their expressions now. We start by considering
\begin{equation}
\begin{aligned}
kG^{sl}_{-\frac{1}{2}}\left[\ket{m,j}^{NS}\right]^{\sigma^w}=k\left[\sigma^w\left(G^{sl}_{-\frac{1}{2}}\right)\ket{m,j}^{NS}\right]^{\sigma^w}\,.
\end{aligned}
\end{equation}
Using \eqref{eq:ade3G}, we obtain
\begin{equation}
\begin{aligned}
k\left[\sigma^w\left(G^{sl}_{-\frac{1}{2}}\right)\ket{m,j}^{NS}\right]^{\sigma^w}=&\left[\sigma^w\left(\sum_n\left( 
\mathcal{J}^+_n\psi^-_{-\frac{1}{2}-n}+\mathcal{J}^-_n\psi^+_{-\frac{1}{2}-n}-2\mathcal{J}^3_n\psi^3_{-\frac{1}{2}-n}\right)\right.\right.\\
&\hspace{2cm}\left.\left.-\sum_{r+s+p=-\frac{1}{2}}\frac{2}{k}:\psi^+_r\psi^-_s\psi^3_p: \right)\ket{m,j}^{NS}\right]^{\sigma^w}\\
=&\left[\left(\sum_n\left( 
\mathcal{J}^+_{n-w}\psi^-_{-\frac{1}{2}-n+w}+\mathcal{J}^-_{n+w}\psi^+_{-\frac{1}{2}-n-w}\right)\right.\right.\\
&\hspace{0.75cm}-2\sum_n\left(\mathcal{J}^3_n+\frac{(k+2)w}{2}\delta_{n,0}\right)\psi^3_{-\frac{1}{2}-n}\\
&\hspace{1.25cm}\left.\left.-\sum_{r+s=0}\frac{2}{k}:\psi^+_{r-w}\psi^-_{s+w}\psi^3_{-\frac{1}{2}}:+2w\psi^3_{-\frac{1}{2}} \right)\ket{m,j}^{NS}\right]^{\sigma^w}\,.
\label{eq:Gdesccomputation}
\end{aligned}
\end{equation}
In writing the fermion cubic term, we have imposed the creation-annihilation normal ordering $:...:$. This means that we put the lowering operators to the right and the raising operators to the left. For the first three terms in the second equality of \eqref{eq:Gdesccomputation}, only the zero modes of the currents contribute since otherwise the currents or the fermions annihilate the ground state. As for the last term in the first equality of \eqref{eq:Gdesccomputation}, only $p=-\tfrac{1}{2}$ contributes by a similar reasoning. This contribution can be computed by noting that the shift in the mode numbers destroy the normal ordering and in order to rescue this back, one has to re-order precisely $w$ terms which gives the last line of \eqref{eq:Gdesccomputation}. Hence, from the calculation above, we obtain
\begin{equation}
\begin{aligned}
G^{sl}_{-\frac{1}{2}}\left[\ket{m,j}^{NS}\right]^{\sigma^w}=\frac{(m+j)}{k}V^w_{m+1,j}\delta_{w+1}(\psi^-)+(m-j)V^w_{m-1,j}\delta_{w-1}(\psi^-)&\\
-\frac{2h}{k}V^w_{m,j}\psi^3\delta_{w}(\psi^-)&\,,
\label{eq:Gdescendant}
\end{aligned}
\end{equation}
where we have defined 
\begin{equation}
h:=m+\frac{kw}{2}\,,
\end{equation}
which is the eigenvalue of $\bigl[\ket{m,j}^{NS}\bigr]^{\sigma^w}$ under $J^3_0$. Holographically, $h$ will be identified with the conformal weight in the dual CFT.

\subsection[The \texorpdfstring{$x$}{x} basis]{\boldmath The \texorpdfstring{$x$}{x} basis}
In order to facilitate the holographic interpretation, it is useful to consider vertex operators in the so-called \emph{$x$ basis}. This is obtained by conjugating vertex operators \eqref{eq:general vertex operator NS sector}, \eqref{eq:general vertex operator R sector}, \eqref{eq:Gdescendant} with $e^{xJ^+_0}$. Note that we use the current $J^+_0$ in the definition of the $x$ basis vertex operators. It turns out that this is the right current to use instead of the decoupled analogue $\mathcal{J}^+_0$ as we will see from the subsequent calculations in Section \ref{sec:correlator}. Since the vertex operators we are primarily interested in is a tensor product of bosonic $\mathfrak{sl}(2,\RR)_{k+2}$ operator and the fermions, and that the operator $e^{xJ^+_0}$ factorises into two commuting parts 
\begin{equation}
e^{xJ^+_0}=e^{x\mathcal{J}^+_0}e^{-\frac{2x}{k}(\psi^3\psi^+)_0}\,,
\end{equation}
we can consider the effects of the conjugation separately.
\subsubsection*{NS sector}
For the bosonic current theory, we simply have
\begin{equation}
\begin{aligned}
e^{x\beta_0}\Phi e^{-x\beta_0}=\Phi\,,\quad e^{x\beta_0}\gamma e^{-x\beta_0}=\gamma-x\,,\quad e^{x\beta_0}\beta e^{-x\beta_0}=\beta\,,
\end{aligned}
\end{equation}
where we have used that $\mathcal{J}^+_0=\beta_0$ as implied by \eqref{eq:defWaki} whereas for the fermion theory, we have
\begin{equation}
\begin{aligned}
e^{-\frac{2x}{k}(\psi^3\psi^+)_0} \psi^+ e^{\frac{2x}{k}(\psi^3\psi^+)_0}=&\psi^+\,,\\
e^{-\frac{2x}{k}(\psi^3\psi^+)_0} \psi^3 e^{\frac{2x}{k}(\psi^3\psi^+)_0}=&\psi^3-x\psi^+\,,\\
e^{-\frac{2x}{k}(\psi^3\psi^+)_0} \psi^- e^{\frac{2x}{k}(\psi^3\psi^+)_0}=&\psi^--2x\psi^3+x^2\psi^+\,.
\end{aligned}
\end{equation}
Thus, a general NS sector vertex operator \eqref{eq:general vertex operator NS sector} in the $x$ basis is given by
\begin{equation}
\begin{aligned}
V\left( \left[\ket{m,j}^{NS}\right]^{\sigma^w},z,x\right)=&e^{(w/Q-Qj)\Phi}\bigg(\frac{\partial^w\gamma}{w!}\bigg)^{-m-j}\delta_w(\gamma-x)\delta_w(\psi^--2x\psi^3+x^2\psi^+)\\
=&e^{(w/Q-Qj)\Phi}\bigg(\frac{\partial^w\gamma}{w!}\bigg)^{-m-j}\delta_w(\gamma-x)\delta_w(\psi^-,x)\,,
\label{eq:x-basis general vertex operator}
\end{aligned}
\end{equation}
where we have defined
\begin{equation}
\delta_w(\psi^-,x):=\delta_w(\psi^--2x\psi^3+x^2\psi^+)\,,
\end{equation}
for notational convenience.
The $x$ basis for the supercurrent descendant \eqref{eq:Gdescendant} is
\begin{equation}
\begin{aligned}
&V\left( \left[G^{sl}_{-\frac{1}{2}}\ket{m,j}^{NS}\right]^{\sigma^w},z,x\right)\\
&=\frac{(m+j)}{k}e^{(w/Q-Qj)\Phi}\bigg(\frac{\partial^w\gamma}{w!}\bigg)^{-m-j-1}\delta_w(\gamma-x)\delta_{w+1}(\psi^-,x)\\
&\hspace{1cm}+(m-j)e^{(w/Q-Qj)\Phi}\bigg(\frac{\partial^w\gamma}{w!}\bigg)^{-m-j+1}\delta_w(\gamma-x)\delta_{w-1}(\psi^-,x)\\
&\hspace{2cm}-\frac{2h}{k}e^{(w/Q-Qj)\Phi}\bigg(\frac{\partial^w\gamma}{w!}\bigg)^{-m-j}\delta_w(\gamma-x)(\psi^3-x\psi^+)\delta_{w}(\psi^-,x)\,.
\label{eq:x-basis G descendant}
\end{aligned}
\end{equation}

\subsubsection*{R sector}
In the R sector, we conjugate the fermionic part in \eqref{eq:general vertex operator R sector} by $e^{-\frac{2x}{k}(\psi^3\psi^+)_0}$. In the computations we will do later on, only the $x$ basis of $e^{-\frac{iH}{2}}V(\ket{0}_{R,\text{rest}})$ matters and the result is given by
\begin{equation}
e^{-\frac{2x}{k}(\psi^3\psi^+)_0}e^{-\frac{iH}{2}}V(\ket{0}_{R,\text{rest}})e^{\frac{2x}{k}(\psi^3\psi^+)_0}=e^{-\frac{iH}{2}}V(\ket{0}_{R,\text{rest}})-\frac{2ix}{\sqrt{k}}e^{\frac{iH}{2}}V(\psi^3_0\ket{0}_{R,\text{rest}})\,.
\label{eq:x basis R sector}
\end{equation}

\subsection{Screening operator}
Eventually, $\gamma$ will be interpreted as a covering map which is a map from the worldsheet $\Sigma$ to the boundary sphere of $\rm AdS_3$, subject also to some conditions to be discussed shortly. Obviously, this map will have poles at the preimages of infinity. However, $\gamma$ as a free field cannot have poles if there is nothing inserted at those points. The solution to this problem is to add a \emph{screening operator} $\mathcal{O}$\footnote{The relation between this screening operator and the one appearing in \cite{Eberhardt:2019ywk} as well as the $W$ field in \cite{Dei:2020zui} is discussed in \cite{Knighton:2023mhq,Knighton:2024qxd}.} to the action \eqref{eq:ads full action} \cite{Knighton:2023mhq,Knighton:2024qxd,Dei:2023ivl}. In addition to inducing poles in $\gamma$, this operator should not destroy the existing symmetry of the $\mathfrak{sl}(2,\RR)$ theory, in particular, it should be invisible to the $\mathfrak{sl}(2,\RR)_{k+2}$ currents. This suggests that $\mathcal{O}$ should be singlet under $SL(2,\mathbb{R})$ symmetry. The screening operator takes the form
\begin{equation}
\mathcal{O}=\int_\Sigma D(z)\bar D(\bar z)\,.
\end{equation}
This requires the worldsheet conformal weight of $D$ ($\bar D$) to be (1,0) ((0,1)) in order for $\mathcal{O}$ to be marginal on the worldsheet. Such $D$ has been found for the bosonic string theory in $\text{AdS}_3\times X$ in the near-boundary limit \cite{Knighton:2023mhq,Knighton:2024qxd} in which case $D$ is given by
\begin{equation}
D(z)=e^{-2\Phi/Q}\left(\oint_z\frac{\mathrm{d}y}{2\pi i}\,\gamma(y)\right)^{-(k_b-1)}\delta(\beta)\,.
\end{equation}
The properties described above can be shown to be satisfied. However since we are now in superstring theory, we have to further ensure that this $D$ we have found is supersymmetric or find a supersymmetric completion of it. We will now show that $D$ itself is already supersymmetric, that is, $D$ is a supercurrent descendant.

Consider the field 
\begin{equation}
\varphi:=V\left( \left[\Ket{\frac{k}{2},\frac{k}{2}}^{NS}\right]^{\sigma^{-1}} ,z\right)=V^{-1}_{\frac{k}{2},\frac{k}{2}}\psi^+\,.
\end{equation}
This state has worldsheet conformal weight $\tfrac{1}{2}$ since $\psi^+$ has conformal weight $\tfrac{1}{2}$ and the conformal weight of the bosonic part is
\begin{equation}
\Delta=\frac{j(1-j)}{k}-\left(m+\frac{(k+2)w}{2}\right)w+\frac{(k+2)w^2}{4}=\frac{(k/2)(1-k/2)}{k}-1+\frac{(k+2)}{4}=0\,.
\end{equation}
Next, by considering the $G^{sl}$ descendant of this state and using \eqref{eq:Gdescendant}, we obtain
\begin{equation}
\begin{aligned}
G^{sl}_{-\frac{1}{2}}\cdot\varphi=V\left(G^{sl}_{-\frac{1}{2}}\left[\Ket{\frac{k}{2},\frac{k}{2}}^{NS}\right]^{\sigma^{-1}},z\right)=V^{-1}_{\frac{k+2}{2},\frac{k}{2}}=e^{-2\Phi/Q}\left(\oint_z\frac{\mathrm{d}y}{2\pi i}\,\gamma(y)\right)^{-(k+1)}\delta(\beta)\,,
\label{}
\end{aligned}
\end{equation}
where we have used $k_b=k+2$ in the above expression.
The second term in \eqref{eq:Gdescendant} vanishes since $m=j=\tfrac{k}{2}$ for $\varphi$ and the third term vanishes since $h=m+\tfrac{kw}{2}=\tfrac{k}{2}-\tfrac{k}{2}=0$. Thus, we see that the $G^{sl}$ descendant of the state $\varphi$ is precisely $D$ found in the bosonic theory \cite{Knighton:2024qxd,Knighton:2023mhq}. Note that this descendent is purely bosonic and thus is automatically invisible to the fermions. Also, the screening operator $\mathcal{O}$ is a singlet under $SL(2,\RR)$ by exactly the same bosonic argument in \cite{Knighton:2023mhq}. This implies that the bosonic screening operator $D$ is the right screening operator even for superstring consideration. 

\subsection{Correlation functions}\label{sec:bosonic correlation functions}
The action in the near-boundary limit, with the inclusion of the screening operator, is given by 
\begin{equation}
\begin{aligned}
S_{\rm AdS_3}=\frac{1}{2\pi}\int\left(\frac{1}{2}\partial\Phi\,\overline{\partial}\Phi+\beta\overline{\partial}\gamma+\bar{\beta}\partial\bar{\gamma}-\frac{Q}{4}R\Phi-2\pi pD\right)+S_{\text{free fermions}}\,,
\label{eq:action}
\end{aligned}
\end{equation}
in the superstring case. The correlators we are interested in are the ones with insertions of \eqref{eq:x-basis general vertex operator}, \eqref{eq:x-basis G descendant}, \eqref{eq:x basis R sector}. In this subsection we do a warm up by reviewing how correlators are calculated in the purely bosonic theory, namely, in the theory described by the first term in \eqref{eq:action}. The full calculation is rather involved and we will only mention important points. To simplify our presentation, we will be focusing only on the left movers and we will mostly drop numerical constants. We will, however, try to make it clear which equations are actual equalities ($=$) and which are equalities up to some constant ($\simeq$).

In the bosonic string theory, the sphere correlators we interested in are
\begin{equation}
\begin{aligned}
\Braket{V^{w_1}_{m_1,j_1}(z_1,x_1)...V^{w_n}_{m_n,j_n}(z_n,x_n)}\Braket{c(z_1)c(z_2)c(z_3)}\braket{X}\,,
\end{aligned}
\end{equation}
where $V^w_{m,j}$ is the bosonic part in \eqref{eq:x-basis general vertex operator}. We have also included the compact CFT $X$ explicitly and $\braket{X}$ denotes collectively the correlator in the compact $X$ theory. The ghost contribution gives 
\begin{equation}
\Braket{c(z_1)c(z_2)c(z_3)}=z_{12}z_{13}z_{23}\,,
\end{equation}
where we define 
\begin{equation}
z_{ij}:=z_i-z_j\,.
\end{equation}
The correlation function involving $V^w_{m,j}$'s is computed in \cite{Knighton:2024qxd} using the first term in \eqref{eq:action}. In the full $SL(2,\RR)$ WZW model, the correlation function $\braket{V^{w_1}_{m_1,j_1}...V^{w_n}_{m_n,j_n}}$ exhibits divergences when the sum of $\mathfrak{sl}(2,\RR)$ spins $j_i$ takes specific values. In the near-boundary approximation, this correlation function vanishes unless the $\Phi$ momentum conservation is satisfied. This correlation function thus contains a (radial) $\Phi$ momentum conservation delta function (whose support is the position of the pole in the full correlator) and the coefficient of the delta function is, up to some normalisation, the residue of the full correlator. Hence, the residue can be computed in the near-boundary limit and is given by
\begin{equation}
\begin{aligned}
\Braket{V^{w_1}_{m_1,j_1}...V^{w_n}_{m_n,j_n}}'\simeq\frac{1}{z_{12}z_{13}z_{23}}\times(C^{\gamma})^{\frac{k_{b}}{2}}\prod_{a=1}^{N}(c_a^{})^{-\frac{k_{b}}{2}}\prod_{i=1}^{n}(a_i^{})^{-h_i+\frac{k_{b}(w_i-1)}{4}}&\\
\prod_{i<j}(z_i-z_j)^{-(Qj_i-1/Q)(Qj_j-1/Q)}\delta^{(n-3)}(\gamma-\gamma^*)&\,,
\label{eq:boson WZW without the delta function}
\end{aligned}
\end{equation}
if the condition 
\begin{equation}
-Q^2\left( \sum_ij_i-\frac{k_b}{2}(n-2)+(n-3) \right)=0\,,
\label{eq:j constraint}
\end{equation}
holds. We put a prime on the correlator to emphasise that we have dropped the momentum conservation delta function. Later, we will drop the prime but it should be understood that we implicitly ignore the delta function. This condition \eqref{eq:j constraint} is referred to as the \emph{j-constraint} \cite{Eberhardt:2019ywk,Knighton:2024qxd}. Note that there is another kind of delta function in \eqref{eq:boson WZW without the delta function} which we write as $\delta^{(n-3)}(\gamma-\gamma^*)$. This is the delta function that localises worldsheet amplitudes to isolated points in the worldsheet moduli space $\mathcal{M}_{0,n}$ and its normalisation is chosen such that 
\begin{equation}
\int_{\mathcal{M}_{0,n}}F(\gamma)\delta^{(n-3)}(\gamma-\gamma^*)=\sum_{\gamma} F(\gamma)\,.
\end{equation}
In particular, at the locus of the localisation, the $\gamma$ field has to satisfy
\begin{equation}
\gamma(z)=x_i+O((z-z_i)^{w_i}),\text{ as }z\to z_i\,,
\end{equation}
for all $i$. We will refer to $\gamma$ that satisfies the above condition as a (branched) \emph{covering map}.
In writing the result above, we have defined $c_a$'s as the residues at the poles of $\gamma$ while
\begin{equation}
a_i:=\frac{\partial^{w_i}\gamma(z_i)}{w_i!}\,
\label{eq:definition of a_i}
\end{equation}
and $C^\gamma$ is defined by
\begin{equation}
\partial\gamma=C^\gamma\frac{\prod_{i=1}^n(z-z_i)^{w_i-1}}{\prod_{a=1}^N(z-\lambda_a)^2}\,,
\label{eq:definition of C^gamma}
\end{equation}
where $\lambda_a$ denote the poles of $\gamma$. The number of poles $N$ is given in terms of the branching indices $w_i$ by 
\begin{equation}
N=1+\sum_{i=1}^{n}\frac{w_i-1}{2}\,.
\label{eq:degree with j constraint}
\end{equation}
Putting everything together and integrating over the worldsheet moduli, we arrive at
\begin{equation}
\begin{aligned}
&\int_{\mathcal{M}_{0,n}}\Braket{V^{w_1}_{m_1,j_1}...V^{w_n}_{m_n,j_n}}\Braket{c(z_1)c(z_2)c(z_3)}\braket{X}\\
&\simeq\sum_{\gamma}(C^{\gamma})^{\frac{k_b}{2}}\prod_{a=1}^{N}(c_a^{})^{-\frac{k_b}{2}}\prod_{i=1}^{n}(a_i^{})^{-h_i+\frac{k_b(w_i-1)}{4}}\left(\prod_{i<j}(z_i-z_j)^{-(Qj_i-1/Q)(Qj_j-1/Q)}\braket{X}\right)\,.
\label{eq:bosonic wsamplitude with j constraint}
\end{aligned}
\end{equation}
This is precisely the form of a correlator in a symmetric product orbifold with a seed theory being $\RR_{\mathcal{Q}}\times X$ \cite{Knighton:2024qxd,Dei:2019iym} with $\mathcal{Q}$ given by \eqref{eq:dilaton slope}. 

If the $j$-constraint \eqref{eq:j constraint} is not fulfilled, we have instead of \eqref{eq:j constraint}
\begin{equation}
m=-Q^2\left( \sum_ij_i-\frac{k_b}{2}(n-2)+(n-3) \right)\,,
\label{eq:generalised j constraint}
\end{equation}
with $m$ being a positive integer. Also, the number of poles is modified from \eqref{eq:degree with j constraint} to
\begin{equation}
N=1+\sum_{i=1}^{n}\frac{w_i-1}{2}+\frac{m}{2}\,.
\label{eq:degree without j constraint}
\end{equation}
In this case, the worldsheet amplitude takes the following form
\begin{equation}
\begin{aligned}
&\int_{\mathcal{M}_{0,n}}\Braket{V^{w_1}_{m_1,j_1}...V^{w_n}_{m_n,j_n}}\Braket{c(z_1)c(z_2)c(z_3)}\braket{X}\\
&\simeq\int \prod_{l=1}^md\gamma(\zeta_l)\sum_{\gamma}(C^{\gamma})^{\frac{6k_b}{12}}\prod_{a=1}^{N}(c_a)^{-\frac{6k_b}{12}}\prod_{i=1}^{n}(a_i^{})^{-h_i+\frac{6k_b(w_i-1)}{24}}\prod_{\ell=1}^{m}(\tilde a_\ell^{})^{-1+\frac{6k_b}{24}}\\
&\hspace{2cm}\times\left(\prod_{i<j}(z_i-z_j)^{-\alpha_i\alpha_j}\prod_{i,\ell}(z_i-\zeta_{\ell})^{-\alpha_i\alpha}\prod_{\ell<k}(\zeta_\ell-\zeta_k)^{-\alpha^2}\braket{X}\right)\,,
\label{eq:bosonic wsamplitude without j constraint}
\end{aligned}
\end{equation}
where we have defined 
\begin{equation}
\alpha:=\frac{1}{Q},\quad \alpha_i:=Qj_i-\frac{1}{Q},\quad \tilde a_\ell:=\frac{\partial^2\gamma(\zeta_\ell)}{2}\,.
\label{eq:definition of alpha}
\end{equation}
Here, $\zeta_\ell$'s are the positions of the extra branched point satisfying 
\begin{equation}
\gamma(z)=\gamma(\zeta_\ell)+O((z-\zeta_\ell)^2)\,.
\end{equation}
This worldsheet amplitude \eqref{eq:bosonic wsamplitude without j constraint} is precisely the $m-$th order perturbation in the deformed CFT 
\begin{equation}
\text{Sym}^\infty(\RR_{\mathcal{Q}}\times X)+\int \sigma_{2,\alpha}
\end{equation}
with $\sigma_{2,\alpha}$ denotes a twist-2 field carrying dilaton momentum $\alpha$. Hence, we see that in the bosonic string theory, the near-boundary approximation is sufficiently powerful to allow one to calculate $n$-point worldsheet amplitudes. Furthermore, one can read off immediately the dual CFT. We will shortly see that the inclusion of worldsheet fermions increases the complication of the computation. Nevertherless, it is the fermions that lead us to a sensible holography.

\section{Amplitude computations}\label{sec:correlator}
We have done all the preparations in the previous sections, in this section we will compute some worldsheet $3$-point and $4$-point amplitudes and show that they reproduce correlators in a deformed symmetric product orbifold which we will determine. We will divide the computations into two main categories: those correlators that satisfy the $j$-constraint \eqref{eq:j constraint} and those that satisfy a more general condition \eqref{eq:generalised j constraint}. This distinction is only for presentation and calculational convenience. 

\subsection[\texorpdfstring{Amplitudes satisfying the $j$-constraint}{Amplitudes satisfying the j-constraint}]{\boldmath \texorpdfstring{Amplitudes satisfying the $j$-constraint}{Amplitudes satisfying the j-constraint}}
\subsubsection*{\boldmath \texorpdfstring{Case I: $w=(1,1,1)$ NS sector}{Case I: w=(1,1,1) NS sector}}\label{sec:Case I w=(1,1,1)}
Let us start with the simplest correlator, which is a $3$-point function of spectrally flowed ground states with one spectral flow unit. The correlator we are interested in takes the form
\begin{equation}
\begin{aligned}
\Braket{\left( c(z_1)e^{-\phi(z_1)}V^{1,NS}_{m_1,j_1}(z_1,x_1) \right)\left( c(z_2)e^{-\phi(z_2)}V^{1,NS}_{m_2,j_2}(z_2,x_2) \right)\left( c(z_3)G^{sl}_{-\frac{1}{2}}\cdot V^{1,NS}_{m_3,j_3}(z_3,x_3) \right)}\braket{X}
\label{eq:starting point for NS w=(1,1,1)}
\end{aligned}
\end{equation}
where we display all the ghosts, $\mathfrak{sl}(2,\RR)_{k+2}$ and the $X$ factors explicitly. To save some writing later on, we will refer to the collective $\mathfrak{sl}(2,\RR)_{k+2}$ and $c$ ghost contributions as the `bosonic' contribution, unless stated otherwise. We will also often suppress the explicit ghost and compact CFT $X$ dependence. Notice that even though we strictly should use $G=G^{sl}+G^X$ in the third insertion in \eqref{eq:starting point for NS w=(1,1,1)}, the vertex operator with the action of $G^X$ contains one fermion and thus, the correlator \eqref{eq:starting point for NS w=(1,1,1)} with the action of $G^X$ has three fermions in total and is therefore vanishing. 

The $\phi$ ghost contribution can be computed easily and we get
\begin{equation}
\begin{aligned}
\Braket{e^{-\phi(z_1)}e^{-\phi(z_2)}}=&\frac{1}{z_{12}}\,.
\end{aligned}
\end{equation}
Now, let us consider the $\mathfrak{sl}(2,\RR)^{(1)}_k$ correlator. Using equations \eqref{eq:x-basis general vertex operator} and \eqref{eq:x-basis G descendant}, the correlator is a sum of three terms
\begin{equation}
\begin{aligned}
\frac{m_3+j_3}{k}\langle V^1_{m_1,j_1}(z_1,x_1)V^1_{m_2,j_2}(z_2,x_2)V^1_{m_3+1,j_3}(z_3,x_3)\rangle\langle\delta_1(\psi^-,x_1)\delta_1(\psi^-,x_2)\delta_2(\psi^-,x_3)\rangle&\\
+(m_3-j_3)\langle V^1_{m_1,j_1}(z_1,x_1)V^1_{m_2,j_2}(z_2,x_2)V^1_{m_3-1,j_3}(z_3,x_3)\rangle\langle\delta_1(\psi^-,x_1)\delta_1(\psi^-,x_2)\rangle&\\
-\frac{2h_3}{k}\langle V^1_{m_1,j_1}(z_1,x_1)V^1_{m_2,j_2}(z_2,x_2)V^1_{m_3,j_3}(z_3,x_3)\rangle\hspace{1cm}&\\
\hspace{1cm}\times\langle\delta_1(\psi^-,x_1)\delta_1(\psi^-,x_2)(\psi^3-x_3\psi^+)\delta_1(\psi^-,x_3)\rangle&\,.
\label{eq:full sl(2,R) w=(1,1,1)}
\end{aligned}
\end{equation}
Each term factorises into the bosonic current part and the fermion part. The bosonic contribution (bosonic current $+ c$ ghost) can be computed as reviewed in Section \ref{sec:bosonic correlation functions}. So, let us now compute fermion correlators. The fermion part of the first term in \eqref{eq:full sl(2,R) w=(1,1,1)} is
\begin{equation}
\begin{aligned}
\langle\delta_1(\psi^-,x_1)\delta_1(\psi^-,x_2)\delta_2(\psi^-,x_3)\rangle\,.
\end{aligned}
\end{equation}
Using \eqref{eq:refermionising}, we can write the correlator explicitly as
\begin{equation}
\begin{aligned}
\left\langle(\psi^--2x_1\psi^3+x_1^2\psi^+)(\psi^--2x_2\psi^3+x_2^2\psi^+)\prod_{i=0}^1\partial^i_3(\psi^--2x_3\psi^3+x_3^2\psi^+)\right\rangle\,.
\end{aligned}
\end{equation}
This can be computed using Wick contractions and the result is
\begin{equation}
\frac{k^2z_{12}x_{13}^2x_{23}^2}{z_{13}^2z_{23}^2}\,.
\label{eq:first fermionic w=(1,1,1)}
\end{equation}
The second fermion contribution in \eqref{eq:full sl(2,R) w=(1,1,1)} is
\begin{equation}
\langle(\psi^--2x_1\psi^3+x_1^2\psi^+)(\psi^--2x_2\psi^3+x_2^2\psi^+)\rangle\,,
\end{equation}
which gives
\begin{equation}
\frac{kx_{12}^2}{z_{12}}\,.
\end{equation}
Finally, the last fermion contribution is
\begin{equation}
\langle(\psi^--2x_1\psi^3+x_1^2\psi^+)(\psi^--2x_2\psi^3+x_2^2\psi^+)(\psi^3-x_3\psi^+)(\psi^--2x_3\psi^3+x_3^2\psi^+)\rangle\,,
\end{equation}
which gives
\begin{equation}
\frac{k^2x_{12}x_{13}x_{23}}{z_{13}z_{23}}\,.
\end{equation}
Note that from the localisation in the bosonic $\mathfrak{sl}(2,\RR)$ part (see the discussion in Section \ref{sec:bosonic correlation functions}), the $\gamma$ field satisfies
\begin{equation}
\begin{aligned}
\gamma(z_1)=x_1,\quad\gamma(z_2)=x_2,\quad\gamma(z_3)=x_3\,,
\end{aligned}
\end{equation}
and from \eqref{eq:degree with j constraint}, $\gamma$ has a simple pole and therefore has an expansion 
\begin{equation}
\gamma(z)=b+\frac{c}{z-\lambda}\,.
\end{equation}
We thus see that in this case, we have
\begin{equation}
\begin{aligned}
x_{ij}=-\frac{c z_{ij}}{(z_i-\lambda)(z_j-\lambda)},\quad C^\gamma=-c,\quad a_i=-\frac{c}{(z_i-\lambda)^2}\,.
\label{eq:covering map data w=(1,1,1)}
\end{aligned}
\end{equation}
Hence, we can rewrite the fermion contributions as
\begin{equation}
\begin{aligned}
\frac{k^2z_{12}x_{13}^2x_{23}^2}{z_{13}^2z_{23}^2}=&k^2z_{12}a_1a_2(a_3)^2\,,\\
\frac{kx_{12}^2}{z_{12}}=&kz_{12}a_1a_2\,,\\
\frac{k^2x_{12}x_{13}x_{23}}{z_{13}z_{23}}=&k^2z_{12}a_1a_2a_3\,.
\end{aligned}
\end{equation}
Combining with the bosonic and $\phi$ contributions, we obtain
\begin{equation}
\begin{aligned}
&\sum_\gamma\frac{1}{z_{12}}\times(C^{\gamma})^{\frac{k+2}{2}}(c)^{-\frac{k+2}{2}}\prod_{i=1}^{3}(a_i)^{-(m_i+\frac{(k+2)}{2})}\prod_{i<j}(z_i-z_j)^{-(Qj_i-1/Q)(Qj_j-1/Q)}\\
&\times\left(\frac{m_3+j_3}{k}\times\frac{1}{a_3}\times k^2z_{12}a_1a_2(a_3)^2+(m_3-j_3)\times a_3\times kz_{12}a_1a_2-\frac{2h_3}{k}\times k^2z_{12}a_1a_2a_3\right)\\
&\simeq\sum_\gamma(C^{\gamma})^{\frac{k}{2}}(c)^{-\frac{k}{2}}\prod_{i=1}^{3}(a_i)^{-\left(m_i+\frac{k}{2}\right)}\prod_{i<j}(z_i-z_j)^{-(Qj_i-1/Q)(Qj_j-1/Q)}\\
&\simeq\sum_\gamma(C^{\gamma})^{\frac{k}{2}}(c)^{-\frac{k}{2}}\prod_{i=1}^{3}(a_i^{})^{-h_i}\prod_{i<j}(z_i-z_j)^{-(Qj_i-1/Q)(Qj_j-1/Q)}\,.
\end{aligned}
\end{equation}
In order to go to the third line, we have used \eqref{eq:covering map data w=(1,1,1)}. Recovering the compact CFT correlator $\braket{X}$ that we have suppressed, the worldsheet amplitude in this case is
\begin{equation}
\begin{aligned}
&\simeq\sum_\gamma 
(C^{\gamma})^{\frac{k}{2}}(c)^{-\frac{k}{2}}\prod_{i=1}^{3}(a_i)^{-h_i}\prod_{i<j}(z_i-z_j)^{-(Qj_i-1/Q)(Qj_j-1/Q)}\braket{X}\,.
\end{aligned}
\end{equation}
Before computing another correlator, let us try to extract some information about the dual CFT. First of all, the central charge is $6k$ as can be read off from the exponent of $C^\gamma$ or $c$. Secondly, the spacetime CFT conformal weight is $m_i+\tfrac{kw_i}{2}$ which is precisely the eigenvalue of the coupled current $J^3_0$ as expected. A naive guess for a seed theory is $\RR_{\mathcal{Q}}\times X\times U$ where $\RR_{\mathcal{Q}}$ is a linear dilaton theory, with slope $\mathcal{Q}$ to be determined below, that accounts for the contribution $\prod_{i<j}z_{ij}^{-\alpha_i\alpha_j}$ and $U$ denotes a currently unknown part of the seed theory not captured directly by the correlator we just computed. This is in contrast to the bosonic string theory where we can conclude that $U$ is trivial. This is because the central charge of the dilaton and compact CFT $X$ add up to $6k_b$ in the bosonic setup which saturates the dual CFT central charge in the bosonic consideration. 

Let us now try to infer what $U$ may be. The linear dilaton slope can be computed as in \cite{Knighton:2024qxd} and the result is
\begin{equation}
\mathcal{Q}=Q-\frac{2}{Q}=-\sqrt{\frac{2}{k}}(k-1)\,.
\label{eq:dilaton slope}
\end{equation}
This implies that the dilaton central charge is
\begin{equation}
c_{\text{dilaton}}=1+3\mathcal{Q}^2=6k-11+\frac{6}{k}\,.
\end{equation}
The compact theory $X$ has central charge 
\begin{equation}
15-\frac{3}{2}-\frac{3(k+2)}{k}\,,
\end{equation}
which follows from the criticality of the superstring theory. Hence, the central charge of the factors $\RR_{\mathcal{Q}}\times X$ is $6k-\tfrac{1}{2}$. This implies that the central charge of $U$ is $\tfrac{1}{2}$. If the seed theory has at least an $\mathcal{N}=1$ supersymmetry, this implies that $U$ should be a single free fermion theory. Furthermore, the fermion in $U$ should then be the superpartner of the dilaton in $\RR_{\mathcal{Q}}$.  

To write the seed theory more compactly, we define a notation $\RR_{\mathcal{Q}}^{(1)}:=\RR_{\mathcal{Q}}\times U$. Therefore, we \emph{propose} that a candidate for the dual CFT is a symmetric orbifold theory
\begin{equation}
\text{Sym}^{\infty}\left( \RR_{\mathcal{Q}}^{(1)}\times X \right)\,,
\end{equation}
with possibly some perturbation turned on. We will comment on how this dual CFT fits with various proposals in the literature in Section \ref{sec:discussion}.

\subsubsection*{\boldmath \texorpdfstring{Case II: $w=(1,1,1)$ R sector}{Case II: w=(1,1,1) R sector}}
Let us now look at a slightly more complicated correlator. Since we want to probe the fermionic theory $U$, we should consider states on the worldsheet theory that have a nontrivial fermionic component under the holographic dictionary. This is achieved by considering spectrally flowed vertex operators of the Ramond ground states.

The correlator we now wish to compute is
\begin{equation}
\begin{aligned}
\Braket{\left( c(z_1)e^{-\frac{\phi(z_1)}{2}}V^{1,R}_{m_1,j_1}(z_1,x_1) \right)\left( c(z_2)e^{-\frac{\phi(z_2)}{2}}V^{1,R}_{m_2,j_2}(z_2,x_2) \right)\left( c(z_3)e^{-\phi(z_3)} V^{1,NS}_{m_3,j_3}(z_3,x_3) \right)}\,,
\end{aligned}
\end{equation}
where we have suppressed the compact CFT contribution $\Braket{X}$. The $\phi$ ghost now gives
\begin{equation}
\Braket{e^{-\frac{\phi(z_1)}{2}}e^{-\frac{\phi(z_2)}{2}}e^{-\phi(z_3)}}=\frac{1}{z_{12}^{\frac{1}{4}}z_{13}^{\frac{1}{2}}z_{23}^{\frac{1}{2}}}\,.
\label{eq:phi ghost for R sector}
\end{equation}
Concretely, we pick the fermion part of the correlator to be 
\begin{equation}
\langle V([\psi^-_0\ket{0}_{R,\pm}]^\sigma\otimes\ket{0}_{R,\text{rest}},z_1,x_1)V([\ket{0}_{R,\pm}]^\sigma\otimes\ket{0}_{R,\text{rest}},z_2,x_2)V([\ket{0}]^\sigma,z_3,x_3)\rangle\,.
\end{equation}
Explicitly, the fermion correlator reads\footnote{To satisfy the physical state condition, one may have to dress $\ket{0}_{R,\text{rest}}$ with fermions from $X$. However, since this does not change the conclusion we will draw in this paper, we leave this dressing implicit throughout and only show the $\psi^3_0$ excitations.}
\begin{equation}
\begin{aligned}
\left\langle e^{-\frac{3iH}{2}}(z_1)V(\ket{0}_{R,\text{rest}},z_1)\left(e^{-\frac{iH}{2}}V(\ket{0}_{R,\text{rest}},z_2)+\frac{2ix_{12}}{\sqrt{k}}e^{\frac{iH}{2}}V(\psi^3_0\ket{0}_{R,\text{rest}},z_2)\right)\right.\\
\left.\delta_1(\psi^-,-x_{13})(z_3)\right\rangle\,.
\label{eq:R correlators with w=(1,1,1)}
\end{aligned}
\end{equation}
Note that in \eqref{eq:R correlators with w=(1,1,1)}, we have moved the $x_1-$translation on the first vertex operator to the other two vertex operators at $z_2$ and $z_3$. 

The fermion correlator in \eqref{eq:R correlators with w=(1,1,1)} can be computed by noting that the only contribution comes from the term that has no total $H$ charge. Performing the Wick contractions over $H$ gives (up to a constant)
\begin{equation}
\begin{aligned}
\frac{x_{12}x_{13}^2z_{23}^{\frac{1}{2}}}{z_{12}^{\frac{3}{4}}z_{13}^{\frac{3}{2}}}\left\langle V(\ket{0}_{R,\text{rest}},z_1)V(\psi^3_0\ket{0}_{R,\text{rest}},z_2)\right\rangle\,.
\label{eq:degenerate R sector fermion}
\end{aligned}
\end{equation}
Multiplying by \eqref{eq:phi ghost for R sector} and using \eqref{eq:covering map data w=(1,1,1)}, we get
\begin{equation}
\frac{x_{12}x_{13}^2}{z_{12}z_{13}^2}\left\langle V(\ket{0}_{R,\text{rest}},z_1)V(\psi^3_0\ket{0}_{R,\text{rest}},z_2)\right\rangle\simeq a_1^{\frac{3}{2}}a_2^{\frac{1}{2}}a_3\left\langle V(\ket{0}_{R,\text{rest}},z_1)V(\psi^3_0\ket{0}_{R,\text{rest}},z_2)\right\rangle\,.
\end{equation}
The worldsheet amplitude in this case is
\begin{equation}
\begin{aligned}
&\simeq\sum_{\gamma}(C^{\gamma})^{\frac{k}{2}}(c)^{-\frac{k}{2}}(a_1)^{-\left(m_1+\frac{(k+2)}{2}\right)}(a_2)^{-\left(m_2+\frac{(k+2)}{2}\right)}(a_3)^{-\left(m_3+\frac{(k+2)}{2}\right)}\times a_1^{\frac{3}{2}}a_2^{\frac{1}{2}}a_3\\\
&\hspace{1cm}\times\prod_{i<j}(z_i-z_j)^{-(Qj_i-1/Q)(Qj_j-1/Q)}\Braket{X}_{\text{bos}}\times\left\langle V(\ket{0}_{R,\text{rest}},z_1)V(\psi^3_0\ket{0}_{R,\text{rest}},z_2)\right\rangle\\
&=\sum_{\gamma}(C^{\gamma})^{\frac{k}{2}}(c)^{-\frac{k}{2}}(a_1)^{-\left(m_1-\frac{1}{2}+\frac{k}{2}\right)}(a_2)^{-\left(m_2+\frac{1}{2}+\frac{k}{2}\right)}(a_3)^{-\left(m_3+\frac{k}{2}\right)}\\
&\hspace{1cm}\times\prod_{i<j}(z_i-z_j)^{-(Qj_i-1/Q)(Qj_j-1/Q)}\Braket{X}_{\text{bos}}\times\left\langle V(\ket{0}_{R,\text{rest}},z_1)V(\psi^3_0\ket{0}_{R,\text{rest}},z_2)\right\rangle\,,
\label{eq:final result second R correlator w=(1,1,1)}
\end{aligned}
\end{equation}
where the subscript emphasises that $\braket{X}_{\text{bos}}$ now only contains bosonic contributions from $X$ since we factor out the fermionic contribution explicitly as the last term of \eqref{eq:final result second R correlator w=(1,1,1)}.
We note that the exponent of $a_i$ is precisely the $-J^3_0$ eigenvalue. The last line of \eqref{eq:final result second R correlator w=(1,1,1)} is a correlator in the seed theory $\RR_{\mathcal{Q}}\times U\times X$. As we can see, this involves not just the dilaton and the $X$ theories. We interpret this additional contribution as coming from the fermionic theory $U$. Since $U$ should be the superpartner of the dilaton in $\RR_{\mathcal{Q}}$ and since this dilaton is related to the worldshet dilaton $\Phi$, the fermion in $U$ is likely to be related somewhat simply to the superpartner of $\Phi$. However, we currently do not have a solid evidence for what this fermion precisely is.

Hence, for the cases of $3$-point amplitudes with one unit of spectral flow, we see that we can derive the dual CFT correlators directly from the worldsheet as we have done for the bosonic case \cite{Knighton:2024qxd}. We also see that the holographic dictionary we derived agrees with well-known results: the central charge of the dual seed CFT is $6k$ where $k$ is the level of an $\mathcal{N}=1$ $\mathfrak{sl}(2,\RR)^{(1)}_k$ WZW model and the dual conformal weight is the eigenvalue of $J^3_0$. However, it would be more satisfying if we could calculate higher point amplitudes or amplitudes with higher spectral flow insertions. For the last two examples in this subsection, we shall make our attempts at calculating these types of amplitudes. 

\subsubsection*{\boldmath \texorpdfstring{Case III: $w=(1,1,1,1)$ NS sector}{Case III: w=(1,1,1,1) NS sector}}
We now wish to compute the following amplitude, 
\begin{equation}
\begin{aligned}
\int dz_4\langle V^{1,NS}_{m_1,j_1}(z_1,x_1) V^{1,NS}_{m_2,j_2}(z_2,x_2)\left( G^{sl}_{-\frac{1}{2}}\cdot V^{1,NS}_{m_3,j_3}(z_3,x_3)\right)\left( G^{sl}_{-\frac{1}{2}}\cdot V^{1,NS}_{m_4,j_4}(z_4,x_4)\right)\rangle\,.
\label{eq:sl(2,R) w=(1,1,1,1)}
\end{aligned}
\end{equation}
One may argue that in this case
\begin{equation}
\begin{aligned}
\int dz_4\langle V^{1,NS}_{m_1,j_1}(z_1,x_1) V^{1,NS}_{m_2,j_2}(z_2,x_2)\left( G^{X}_{-\frac{1}{2}}\cdot V^{1,NS}_{m_3,j_3}(z_3,x_3)\right)\left( G^{X}_{-\frac{1}{2}}\cdot V^{1,NS}_{m_4,j_4}(z_4,x_4)\right)\rangle\,,
\label{eq:zero correlator}
\end{aligned}
\end{equation}
contains an even number of fermions and thus may potentially give a nonzero contribution. However, one can show by explicit computations that this correlator is zero as we do in Appendix \ref{sec:tedious calculation in case III}. 

Plugging in the expression for the supercurrent descendant, there are nine terms in \eqref{eq:sl(2,R) w=(1,1,1,1)} to evaluate. However, five of them do not require any new computation since they are just $3$-point functions from the fermion point of view. Those five terms come from
\begin{equation}
\begin{aligned}
&\int dz_4\left\langle V^{1,NS}_{m_1,j_1}(z_1,x_1) V^{1,NS}_{m_2,j_2}(z_2,x_2)\left( G^{sl}_{-\frac{1}{2}}\cdot V^{1,NS}_{m_3,j_3}(z_3,x_3)\right)V^{1,NS}_{m_4-1,j_4}(z_4,x_4) \right\rangle\,,\\
&\int dz_4\left\langle V^{1,NS}_{m_1,j_1}(z_1,x_1) V^{1,NS}_{m_2,j_2}(z_2,x_2)V^{1,NS}_{m_3-1,j_3}(z_3,x_3)\left( G^{sl}_{-\frac{1}{2}}\cdot V^{1,NS}_{m_4,j_4}(z_4,x_4)\right) \right\rangle\,.
\label{eq:correlator with no new computations}
\end{aligned}
\end{equation}
Note that $V^{1,NS}_{m_i-1,j_i}(z_i,x_i)$ comes from the second term in \eqref{eq:Gdescendant}.
Clearly, the first amplitude above gives the same fermionic contribution as in Case I. Hence, we are guaranteed that the exponent of $a_i,i=1,2,3$ will be the eigenvalues of $-J^3_0$ by the computations we have done in that case. Thus, we now need to consider the exponent of $a_4$. Note the shift in $\mathcal{J}^3_0$ eigenvalue of the vertex operator at $z_4$ in the first line of \eqref{eq:correlator with no new computations}. This means that the exponent of $a_4$ from the bosonic contribution is
\begin{equation}
-\left((m_4-1)+\frac{k+2}{2} \right)=-\left( m_4+\frac{k}{2} \right)\,,
\end{equation}
which is already the $-J^3_0$ eigenvalue of the state $\left[\ket{m_4,j_4}^{NS}\right]^{\sigma}$ as desired. The argument for the second amplitude in \eqref{eq:correlator with no new computations} is very similar and we conclude that both amplitudes in \eqref{eq:correlator with no new computations} are of the form
\begin{equation}
\begin{aligned}
\simeq\sum_\gamma 
(C^{\gamma})^{\frac{k}{2}}(c)^{-\frac{k}{2}}\prod_{i=1}^{4}(a_i)^{-h_i}\prod_{i<j}(z_i-z_j)^{-(Qj_i-1/Q)(Qj_j-1/Q)}\braket{X}\,,
\label{eq:expected answer for w=(1,1,1,1)}
\end{aligned}
\end{equation}
and $h_i$ is the $J^3_0$ eigenvalues of $\left[\ket{m_i,j_i}^{NS}\right]^{\sigma}$. 

There are therefore only four correlators we need to compute and verify that they reproduce the expected symmetric orbifold correlator \eqref{eq:expected answer for w=(1,1,1,1)}. The first one is
\begin{equation}
\begin{aligned}
\langle V^1_{m_1,j_1}(z_1,x_1)V^1_{m_2,j_2}(z_2,x_2)V^1_{m_3+1,j_3}(z_3,x_3)V^1_{m_4+1,j_4}(z_4,x_4)\rangle&\\
\times\langle\delta_1(\psi^-,x_1)\delta_1(\psi^-,x_2)\delta_2(\psi^-,x_3)\delta_2(\psi^-,x_4)\rangle&\,.
\label{eq:1st contribution}
\end{aligned}
\end{equation}
The fermion contribution in this case gives
\begin{equation}
\begin{aligned}
\langle\delta_1(\psi^-,x_1)\delta_1(\psi^-,x_2)\delta_2(\psi^-,x_3)\delta_2(\psi^-,x_4)\rangle&=k^3a_1a_2a_3^2a_4^2z_{12}\,.
\end{aligned}
\end{equation}
See also Appendix \ref{sec:tedious calculation in case III} for more detail calculations of this correlator.
Combining this with the bosonic and $\phi$ ghost contributions we see that \eqref{eq:1st contribution} is indeed given by \eqref{eq:expected answer for w=(1,1,1,1)} since
\begin{equation}
\begin{aligned}
\eqref{eq:1st contribution}\simeq&\sum_\gamma (C^{\gamma})^{\frac{k}{2}}(c)^{-\frac{k}{2}}\prod_{i=1}^{4}(a_i)^{-\left(m_i+\frac{kw_i}{2}\right)}\prod_{i<j}(z_i-z_j)^{-(Qj_i-1/Q)(Qj_j-1/Q)}\times(a_1a_2a_3^{2}a_4^{2})^{-1}\\
&\hspace{2cm}\times\frac{1}{z_{12}}\times k^3a_1a_2a_3^2a_4^2z_{12}\\
\simeq&\sum_\gamma (C^{\gamma})^{\frac{k}{2}}(c)^{-\frac{k}{2}}\prod_{i=1}^{4}(a_i)^{-h_i}\prod_{i<j}(z_i-z_j)^{-(Qj_i-1/Q)(Qj_j-1/Q)}\,.
\label{eq:1st contribution from boson and fermion}
\end{aligned}
\end{equation}
The next contribution is
\begin{equation}
\begin{aligned}
&\langle V^1_{m_1,j_1}(z_1,x_1)V^1_{m_2,j_2}(z_2,x_2)V^1_{m_3,j_3}(z_3,x_3)V^1_{m_4+1,j_4}(z_4,x_4)\rangle\\
&\times\langle\delta_1(\psi^-,x_1)\delta_1(\psi^-,x_2)(\psi^3-x_3\psi^+)\delta_1(\psi^-,x_3)\delta_2(\psi^-,x_4)\rangle\,,
\label{eq:2nd and 3rd contribution}
\end{aligned}
\end{equation}
which gives
\begin{equation}
\begin{aligned}
\eqref{eq:2nd and 3rd contribution}\simeq&\sum_\gamma (C^{\gamma})^{\frac{k}{2}}(c)^{-\frac{k}{2}}\prod_{i=1}^{4}(a_i)^{-\left(m_i+\frac{kw_i}{2}\right)}\prod_{i<j}(z_i-z_j)^{-(Qj_i-1/Q)(Qj_j-1/Q)}\times(a_1a_2a_3a_4^{2})^{-1}\\
&\hspace{2cm}\times k^3a_1a_2a_3a_4^2z_{12}\times\frac{1}{z_{12}}\\
\simeq&\sum_\gamma (C^{\gamma})^{\frac{k}{2}}(c)^{-\frac{k}{2}}\prod_{i=1}^{4}(a_i)^{-h_i}\prod_{i<j}(z_i-z_j)^{-(Qj_i-1/Q)(Qj_j-1/Q)}\,,
\label{eq:2nd contribution from boson and fermion}
\end{aligned}
\end{equation}
where the first and second lines of \eqref{eq:2nd contribution from boson and fermion} comes the first and the second lines of \eqref{eq:2nd and 3rd contribution} and the $\phi$ correlator, respectively. The fermion contribution in \eqref{eq:2nd contribution from boson and fermion} is computed in Appendix \ref{sec:tedious calculation in case III}.
The third contribution
\begin{equation}
\begin{aligned}
&\langle V^1_{m_1,j_1}(z_1,x_1)V^1_{m_2,j_2}(z_2,x_2)V^1_{m_3+1,j_3}(z_3,x_3)V^1_{m_4,j_4}(z_4,x_4)\rangle\\
&\times\langle\delta_1(\psi^-,x_1)\delta_1(\psi^-,x_2)\delta_2(\psi^-,x_3)(\psi^3-x_4\psi^+)\delta_1(\psi^-,x_4)\rangle\,.
\end{aligned}
\end{equation}
can be obtained by simply exchanging $3\leftrightarrow4$ in \eqref{eq:2nd and 3rd contribution} which clearly gives us \eqref{eq:expected answer for w=(1,1,1,1)}. Lastly, we compute
\begin{equation}
\begin{aligned}
&\langle V^1_{m_1,j_1}(z_1,x_1)V^1_{m_2,j_2}(z_2,x_2)V^1_{m_3,j_3}(z_3,x_3)V^1_{m_4,j_4}(z_4,x_4)\rangle\\
&\times\langle\delta_1(\psi^-,x_1)\delta_1(\psi^-,x_2)(\psi^3-x_3\psi^+)\delta_1(\psi^-,x_3)(\psi^3-x_4\psi^+)\delta_1(\psi^-,x_4)\rangle\,,
\label{eq:4th contribution}
\end{aligned}
\end{equation}
which gives
\begin{equation}
\begin{aligned}
\eqref{eq:4th contribution}\simeq&\sum_\gamma (C^{\gamma})^{\frac{k}{2}}(c)^{-\frac{k}{2}}\prod_{i=1}^{4}(a_i)^{-\left(m_i+\frac{kw_i}{2}\right)}\prod_{i<j}(z_i-z_j)^{-(Qj_i-1/Q)(Qj_j-1/Q)}\times(a_1a_2a_3a_4)^{-1}\\
&\hspace{2cm}\times\frac{1}{z_{12}}\times k^3a_1a_2a_3a_4z_{12}\\
\simeq&\sum_\gamma (C^{\gamma})^{\frac{k}{2}}(c)^{-\frac{k}{2}}\prod_{i=1}^{4}(a_i)^{-h_i}\prod_{i<j}(z_i-z_j)^{-(Qj_i-1/Q)(Qj_j-1/Q)}\,,
\label{eq:4th contribution from boson and fermion}
\end{aligned}
\end{equation}
where the last term of the second line of \eqref{eq:4th contribution from boson and fermion} is computed in Appendix \ref{sec:tedious calculation in case III}. Re-introducing the $X$ dependence we have suppressed, we conclude that the amplitude \eqref{eq:sl(2,R) w=(1,1,1,1)} is given by
\begin{equation}
\begin{aligned}
\eqref{eq:sl(2,R) w=(1,1,1,1)}=&\sum_\gamma (C^{\gamma})^{\frac{k}{2}}(c)^{-\frac{k}{2}}\prod_{i=1}^{4}(a_i)^{-h_i}\prod_{i<j}(z_i-z_j)^{-(Qj_i-1/Q)(Qj_j-1/Q)}\Braket{X}\,.
\end{aligned}
\end{equation}
One thus sees that $4$-point worldsheet amplitudes (satisfying the $j$-constraint) with $w=(1,1,1,1)$ reproduce correlators in the untwisted sector of the symmetric orbifold ${\rm Sym}^\infty(\RR^{(1)}_{\mathcal{Q}}\times X)$ as expected.

\subsubsection*{\boldmath \texorpdfstring{Case IV: $w=(2,2,1)$ NS sector}{Case IV: w=(2,2,1) NS sector}}
To finish this subsection, let us now compute another correlator to see if our result extends beyond $w=1$ sector. We will now compute the following correlator (with the compact CFT suppressed)
\begin{equation}
\begin{aligned}
\Braket{\left( c(z_1)e^{-\phi(z_1)}V^{2,NS}_{m_1,j_1}(z_1,x_1) \right)\left( c(z_2)e^{-\phi(z_2)}V^{2,NS}_{m_2,j_2}(z_2,x_2) \right)\left( c(z_3)G^{sl}_{-\frac{1}{2}}\cdot V^{1,NS}_{m_3,j_3}(z_3,x_3) \right)}\,,
\end{aligned}
\end{equation}
where we insert two 2-spectrally flowed ground states at $z_1$ and $z_2$ and one $1$-spectrally flowed ground state at $z_3$. Explicitly, the $\mathfrak{sl}(2,\RR)^{(1)}_k$ correlator reads
\begin{equation}
\begin{aligned}
\frac{m_3+j_3}{k}\langle V^2_{m_1,j_1}(z_1,x_1)V^2_{m_2,j_2}(z_2,x_2)V^1_{m_3+1,j_3}(z_3,x_3)\rangle\langle\delta_2(\psi^-,x_1)\delta_2(\psi^-,x_2)\delta_2(\psi^-,x_3)\rangle&\\
+(m_3-j_3)\langle V^2_{m_1,j_1}(z_1,x_1)V^2_{m_2,j_2}(z_2,x_2)V^1_{m_3-1,j_3}(z_3,x_3)\rangle\langle\delta_2(\psi^-,x_1)\delta_2(\psi^-,x_2)\rangle&\\
-\frac{2h_3}{k}\langle V^2_{m_1,j_1}(z_1,x_1)V^2_{m_2,j_2}(z_2,x_2)V^1_{m_3,j_3}(z_3,x_3)\rangle\hspace{1cm}&\\
\hspace{1cm}\times\langle\delta_2(\psi^-,x_1)\delta_2(\psi^-,x_2)(\psi^3-x_3\psi^+)\delta_1(\psi^-,x_3)\rangle&\,.
\label{eq:full sl(2,R) w=(2,2,1)}
\end{aligned}
\end{equation}
The fermion contributions are given in terms of covering map data by
\begin{equation}
\begin{aligned}
\langle\delta_2(\psi^-,x_1)\delta_2(\psi^-,x_2)\delta_2(\psi^-,x_3)\rangle\simeq&\frac{c_1c_2z_{12}(a_1a_2)^{\frac{3}{2}}(a_3)^2}{C^\gamma}\\
\langle\delta_2(\psi^-,x_1)\delta_2(\psi^-,x_2)\rangle\simeq&\frac{c_1c_2z_{12}(a_1a_2)^{\frac{3}{2}}}{C^\gamma}\\
\langle\delta_2(\psi^-,x_1)\delta_2(\psi^-,x_2)(\psi^3-x_3\psi^+)\delta_1(\psi^-,x_3)\rangle\simeq&\frac{c_1c_2z_{12}(a_1a_2)^{\frac{3}{2}}a_3}{C^\gamma}\,.
\end{aligned}
\end{equation}
Computation of these correlators can be found in Appendix \ref{sec:tedious calculation in case IV}.
Taking into account the bosonic and $\phi$ contributions, one obtains
\begin{equation}
\begin{aligned}
&\sum_\gamma(C^{\gamma})^{\frac{k+2}{2}}\prod_{a=1}^2(c_a)^{-\frac{k+2}{2}}\prod_{i=1}^{3}(a_i)^{-\left(m_i+\frac{(k+2)w_i}{2}\right)+\frac{(k+2)(w_i-1)}{4}}\prod_{i<j}(z_i-z_j)^{-(Qj_i-1/Q)(Qj_j-1/Q)}\\
&\hspace{1cm}\times\left(\frac{m_3+j_3}{k}\times\frac{1}{a_3}\times \frac{c_1c_2z_{12}(a_1a_2)^{\frac{3}{2}}(a_3)^2}{C^\gamma}+(m_3-j_3)\times a_3\times \frac{c_1c_2z_{12}(a_1a_2)^{\frac{3}{2}}}{C^\gamma}\right.\\
&\hspace{7cm}\left.-\frac{2h_3}{k}\times \frac{c_1c_2z_{12}(a_1a_2)^{\frac{3}{2}}a_3}{C^\gamma}\right)\times\frac{1}{z_{12}}\times\Braket{X}\\
&\simeq\sum_\gamma (C^{\gamma})^{\frac{k}{2}}\prod_{a=1}^2(c_a)^{-\frac{k}{2}}\prod_{i=1}^{3}(a_i)^{-\left(m_i+\frac{kw_i}{2}\right)+\frac{k(w_i-1)}{4}}\prod_{i<j}(z_i-z_j)^{-(Qj_i-1/Q)(Qj_j-1/Q)}\Braket{X}\\
&\simeq\sum_\gamma (C^{\gamma})^{\frac{k}{2}}\prod_{a=1}^2(c_a)^{-\frac{k}{2}}\prod_{i=1}^{3}(a_i)^{-h_i+\frac{k(w_i-1)}{4}}\prod_{i<j}(z_i-z_j)^{-(Qj_i-1/Q)(Qj_j-1/Q)}\Braket{X}\,,
\end{aligned}
\end{equation}
for the worldsheet amplitude in this case. We see again that the worldsheet $3$-point amplitude with $w=(2,2,1)$ reproduces a symmetric orbifold correlator with the expected central charge and dual CFT conformal weights. We also know from \cite{Knighton:2024qxd} that in order to deduce the information about the dual CFT perturbation, one has to consider those correlators that do not satisfy the $j$-constraint \eqref{eq:generalised j constraint} and this is what we will do next.

\subsection[\texorpdfstring{Amplitudes violating the $j$-constraint}{Amplitudes violating the j-constraint}]{\boldmath \texorpdfstring{Amplitudes violating the $j$-constraint}{Amplitudes violating the j-constraint}}

In this subsection, we explore how allowing for the $j$-constraint to be violated would add to the story we have seen so far. The amplitudes we are going to compute are $3$-point amplitudes with $w=(2,1,1)$ in the NS and R sectors. One of the most important reasons for choosing this amplitude is because the simplest violation of the $j$-constraint \eqref{eq:generalised j constraint} with $m=1$ is possible and this is what we will focus on solely in this subsection. More explicitly, we will assume throughout this subsection that the spin $j_i$'s are chosen such that
\begin{equation}
1=-Q^2\left( \sum_ij_i-\frac{k_b}{2}(n-2)+(n-3) \right)\,.
\label{eq:generalised j constraint with m=1}
\end{equation}
Hence, \eqref{eq:degree without j constraint} becomes
\begin{equation}
\begin{aligned}
N=&1+\sum_i\frac{w_i-1}{2}+\frac{1}{2}\\
=&1+\frac{1}{2}+\frac{1}{2}=2\,.
\label{eq:degree without j constraint and w=(2,1,1) and m=1}
\end{aligned}
\end{equation}
This situation where $m=1$ is dual to correlators in the dual CFT where one inserts a single copy of the perturbation that deforms the dual CFT. Hence, this allows us to extract some information about the perturbation needed to turn on in the dual CFT.

\subsubsection*{\boldmath \texorpdfstring{Case V: $w=(2,1,1)$ NS sector}{Case V: w=(2,1,1) NS sector}}

The correlator we will compute is
\begin{equation}
\begin{aligned}
&\Braket{\left( c(z_1)e^{-\phi(z_1)}V^{2,NS}_{m_1,j_1}(z_1,x_1) \right)\left( c(z_2)e^{-\phi(z_2)}V^{1,NS}_{m_2,j_2}(z_2,x_2) \right)\left( c(z_3) V^{1,NS}_{m_3,j_3}(z_3,x_3) \right)}\braket{X}\\
&=\langle cV^2_{m_1,j_1}(z_1,x_1)cV^1_{m_2,j_2}(z_2,x_2)cV^1_{m_3,j_3}(z_3,x_3)\rangle\times\Braket{V^X(z_1)V^X(z_2)G^X_{-\frac{1}{2}}\cdot V^X(z_3)}\times\frac{1}{z_{12}}\\
&\hspace{8cm}\times\langle\delta_2(\psi^-,x_1)\delta_1(\psi^-,x_2)\delta_1(\psi^-,x_3)\rangle\,.
\label{eq:full sl(2,R) w=(2,1,1)}
\end{aligned}
\end{equation}
Note that it is the piece $G^{X}_{-\frac{1}{2}}\cdot V(z_3)$ that contributes in this case since $G^{sl}_{-\frac{1}{2}}\cdot V(z_3)$ does not have the right number of fermions.
The fermion contribution is
\begin{equation}
\langle\delta_2(\psi^-,x_1)\delta_1(\psi^-,x_2)\delta_1(\psi^-,x_3)\rangle=\frac{k^2x_{12}^2x_{13}^2z_{23}}{z_{12}^2z_{13}^2}\,.
\end{equation}
From \eqref{eq:degree without j constraint and w=(2,1,1) and m=1}, we have $N=2$ and thus the $\gamma$ field takes the form
\begin{equation}
\gamma=b+\frac{c_1}{z-\lambda_1}+\frac{c_2}{z-\lambda_2}\,.
\label{eq:gamma for w=(2,1,1)}
\end{equation}
From this, we obtain using \eqref{eq:definition of C^gamma} and recalling \eqref{eq:definition of a_i}
\begin{equation}
\begin{aligned}
c_1=&C^\gamma\frac{(z_1-\lambda_1)(\zeta-\lambda_1)}{\lambda_{12}^2},&\quad c_2=&C^\gamma\frac{(z_1-\lambda_2)(\zeta-\lambda_2)}{\lambda_{12}^2}\,,\\
2a_1=&\frac{C^\gamma(z_1-\zeta)}{(z_1-\lambda_1)^2(z_1-\lambda_2)^2},&\quad a_2=&\frac{-C^\gamma z_{12}(z_2-\zeta)}{(z_2-\lambda_1)^2(z_2-\lambda_2)^2}\,,\\
a_3=&\frac{-C^\gamma z_{13}(z_3-\zeta)}{(z_3-\lambda_1)^2(z_3-\lambda_2)^2},&\quad 2\tilde a=&\gamma''(\zeta)=\frac{-C^\gamma(z_1-\zeta)}{(\zeta-\lambda_1)^2(\zeta-\lambda_2)^2}\,,\\
-1=&\frac{(z_1-\lambda_1)(\zeta-\lambda_2)}{(\zeta-\lambda_1)(z_1-\lambda_2)}\,,
\label{eq:covering map data for w=(2,1,1)}
\end{aligned}
\end{equation}
where $\zeta$ is the ``extra" branch point, that is,
\begin{equation}
\gamma(z)=\gamma(\zeta)+O((z-\zeta)^2)
\end{equation}
as $z\to\zeta$. Thus, we can write the fermion contribution as
\begin{equation}
\frac{k^2x_{12}^2x_{13}^2z_{23}}{z_{12}^2z_{13}^2}\simeq\frac{z_{12}z_{13}z_{23}(a_1)^{\frac{3}{2}}a_2a_3c_1c_2}{C^\gamma \tilde a^{\frac{1}{2}}(z_1-\zeta)(z_2-\zeta)(z_3-\zeta)}\,.
\end{equation}
Combining with the bosonic \eqref{eq:bosonic wsamplitude without j constraint}, the compact CFT $X$ and ghost contributions, we obtain
\begin{equation}
\begin{aligned}
&\simeq\int d\gamma(\zeta)\sum_{\gamma}(C^{\gamma})^{\frac{6k+12}{12}}\prod_{a=1}^{2}(c_a)^{-\frac{6k+12}{12}}\prod_{i=1}^{3}(a_i^{})^{-\left(m_i+\frac{(k+2)w_i}{2}\right)+\frac{6(k+2)(w_i-1)}{24}}(\tilde a)^{-1+\frac{6k+12}{24}}\\
&\times\left(\prod_{i<j}(z_i-z_j)^{-\alpha_i\alpha_j}\prod_{i}(z_i-\zeta)^{-\alpha_i\alpha}\right)\frac{z_{13}z_{23}a_1^{\frac{3}{2}}a_2a_3c_1c_2\Braket{V^X(z_1)V^X(z_2)(G_{-\frac{1}{2}}\cdot V)^X(z_3)}}{C^\gamma \tilde a^{\frac{1}{2}}(z_1-\zeta)(z_2-\zeta)(z_3-\zeta)}\\
&\simeq\int d\gamma(\zeta)\sum_{\gamma}(C^{\gamma})^{\frac{6k}{12}}\prod_{a=1}^{2}(c_a)^{-\frac{6k}{12}}\prod_{i=1}^{3}(a_i)^{-\left(m_i+\frac{kw_i}{2}\right)+\frac{6k(w_i-1)}{24}}(\tilde a)^{-1+\frac{6k}{24}}\\
&\hspace{1cm}\times\left(\prod_{i<j}(z_i-z_j)^{-\alpha_i\alpha_j}\prod_{i=1}^3(z_i-\zeta)^{-\alpha_i\alpha}\right)\times\frac{z_{13}z_{23}\Braket{V^X(z_1)V^X(z_2)(G_{-\frac{1}{2}}\cdot V)^X(z_3)}}{(z_1-\zeta)(z_2-\zeta)(z_3-\zeta)}\,.
\label{eq:worldsheet result for w=(2,1,1) NS sector}
\end{aligned}
\end{equation}
Recall that $\alpha_i$'s are defined in \eqref{eq:definition of alpha}. From this result, we can read off that the perturbation in the dual CFT has spacetime conformal weight 1 from the exponent of $\tilde a$. Also, this perturbation contains a linear dilaton factor with momentum\footnote{Note that we use the same convention of CFT momentum as \cite{Knighton:2024qxd}. There, they define the field $e^{-q\varphi}$ to have momentum $q$.} 
\begin{equation}
\alpha=\sqrt{\frac{k}{2}}=\frac{1}{Q}\,,
\label{eq:linear dilaton momentum}
\end{equation}
which can be seen from the exponent of $(z_i-\zeta)$ and \eqref{eq:definition of alpha}. The perturbation lives in the $2$-cycle twisted sector. The last term in \eqref{eq:worldsheet result for w=(2,1,1) NS sector} can be used to determine the perturbation in the dual CFT. However, we will not compute this explicitly. This perturbation depends on the detail of the compact CFT $X$ and thus is likely to be model-dependent. Nevertheless, we can say that this perturbation does not depend on the factor $U$ in the seed theory. To see this, we note that all three insertions at $z_i$ lift to the NS ground state as far as the fermion theory $U$ is concerned. Hence, if the perturbation contains nontrivial dependence on $U$, then this will give rise to a sphere $1$-point function in the $U$ theory and is therefore vanishing. Furthermore, because of the appearance of $G^X_{-1/2}$ in \eqref{eq:worldsheet result for w=(2,1,1) NS sector}, we speculate that the perturbation should be a $G^X$ descendant.\footnote{We thank Bob Knighton for a discussion about this point.} We then conclude that the perturbation consists of the linear dilaton factor with momentum $\alpha$ and a ($G^X$-descendant) factor that depends on $X$ and that this perturbation lives in the twist$-2$ sector.

Let us end this subsection by arguing that the appearance of the term 
\begin{equation}
\frac{1}{(z_1-\zeta)(z_2-\zeta)(z_3-\zeta)}\,,
\label{eq:term to give the right conformal weight}
\end{equation}
is expected from the fact that the dual conformal weight of the perturbing field is 1. To extract the conformal weight of the insertion at $\zeta$ (which is the insertion point of the perturbing field), we look at the large $\zeta$ limit of the above expression \eqref{eq:worldsheet result for w=(2,1,1) NS sector} which goes like $\zeta^{-2h}$ with $h$ the conformal weight of the perturbing field. Notice that the term in \eqref{eq:term to give the right conformal weight} contributes $\tfrac{3}{2}$ to the conformal weight of the perturbing field in the seed theory, in addition to the contribution from the linear dilaton factor. Using the symmetric orbifold formula that relates conformal weights in the seed and orbifold theories, we have that the conformal weight in the symmetric orbifold theory is given by
\begin{equation}
h_{\rm sym}=\frac{6k(4-1)}{48}+\frac{1}{2}\left( \frac{\alpha(\mathcal{Q}-\alpha)}{2}+\frac{3}{2} \right)=\frac{3k}{8}+\frac{1}{2}\left(\frac{1}{2} -\frac{3k}{4}+\frac{3}{2} \right)=1\,.
\end{equation}
Here, $\mathcal{Q}$ is the background charge of the dilaton $\RR_{\mathcal{Q}}$ and is given in \eqref{eq:dilaton slope}.

\subsubsection*{\boldmath \texorpdfstring{Case VI: $w=(2,1,1)$ R sector}{Case VI: w=(2,1,1) R sector}}
Let us now compute another amplitude that does not satisfy the $j$-constraint with $m=1$ in \eqref{eq:generalised j constraint}. The amplitude we want to compute is
\begin{equation}
\begin{aligned}
\Braket{\left( c(z_1)e^{-\frac{\phi(z_1)}{2}}V^{2,R}_{m_1,j_1}(z_1,x_1) \right)\left( c(z_2)e^{-\frac{\phi(z_2)}{2}}V^{1,R}_{m_2,j_2}(z_2,x_2) \right)\left( c(z_3)e^{-\phi(z_3)} V^{1,NS}_{m_3,j_3}(z_3,x_3) \right)}\,,
\end{aligned}
\end{equation}
where we pick the fermion correlator to be
\begin{equation}
\left\langle V\left(\left[\ket{0}_{R,\pm}\right]^{\sigma^2}\otimes\ket{0}_{R,\text{rest}},z_1,x_1\right)V\left(\left[\ket{0}_{R,\pm}\right]^\sigma\otimes\ket{0}_{R,\text{rest}},z_2,x_2\right)V(\left[\ket{0}\right]^\sigma,z_3,x_3)\right\rangle\,.
\end{equation}
In fact, this correlator has already been calculated in Case II. The only difference is the covering map data which is now given by \eqref{eq:covering map data for w=(2,1,1)} instead of \eqref{eq:covering map data w=(1,1,1)}. The fermion correlator \eqref{eq:degenerate R sector fermion} and the $\phi$ ghost give
\begin{equation}
\frac{x_{12}x_{13}^2}{z_{12}z_{13}^2}\left\langle V(\ket{0}_{R,\text{rest}},z_1)V(\psi^3_0\ket{0}_{R,\text{rest}},z_2)\right\rangle\,.
\end{equation}
Using \eqref{eq:gamma for w=(2,1,1)} and \eqref{eq:covering map data for w=(2,1,1)} since we also have $N=2$ in this case, we now get
\begin{equation}
\begin{aligned}
\frac{x_{12}x_{13}^2}{z_{12}z_{13}^2}\simeq\frac{c_1c_2a_1(a_2)^{\frac{1}{2}}a_3\tilde a^{-\frac{1}{2}}z_{13}z_{12}^{\frac{1}{2}}}{C^\gamma(z_1-\zeta)^{\frac{3}{2}}(z_2-\zeta)^{\frac{1}{2}}(z_3-\zeta)}\,.
\end{aligned}
\end{equation}
The worldsheet amplitude therefore reads
\begin{equation}
\begin{aligned}
&\simeq\int d\gamma(\zeta)\sum_{\gamma}(C^{\gamma})^{\frac{6k+12}{12}}\prod_{a=1}^{2}(c_a)^{-\frac{6k+12}{12}}\prod_{i=1}^{3}(a_i)^{-\left(m_i+\frac{(k+2)w_i}{2}\right)+\frac{6(k+2)(w_i-1)}{24}}(\tilde a)^{-1+\frac{6k+12}{24}}\\
&\hspace{1cm}\times\left(\prod_{i<j}(z_i-z_j)^{-\alpha_i\alpha_j}\prod_{i}(z_i-\zeta)^{-\alpha_i\alpha}\right)\frac{c_1c_2a_1(a_2)^{\frac{1}{2}}a_3\tilde a^{-\frac{1}{2}}z_{13}z_{12}^{\frac{1}{2}}}{C^\gamma(z_1-\zeta)^{\frac{3}{2}}(z_2-\zeta)^{\frac{1}{2}}(z_3-\zeta)}\\
&\hspace{2cm}\times\left\langle V(\ket{0}_{R,\text{rest}},z_1)V(\psi^3_0\ket{0}_{R,\text{rest}},z_2)\right\rangle\Braket{X}_{\text{bos}}\\
&\simeq\int d\gamma(\zeta)\sum_{\gamma}(C^{\gamma})^{\frac{6k}{12}}\prod_{a=1}^{2}(c_a)^{-\frac{6k}{12}}\prod_{i=1}^{3}(a_i)^{-h_i+\frac{6k(w_i-1)}{24}}(\tilde a)^{-1+\frac{6k}{24}}\\
&\hspace{1cm}\times\left(\prod_{i<j}(z_i-z_j)^{-\alpha_i\alpha_j}\prod_{i=1}^3(z_i-\zeta)^{-\alpha_i\alpha}\right)\times\frac{z_{13}z_{12}^{\frac{1}{2}}}{(z_1-\zeta)^{\frac{3}{2}}(z_2-\zeta)^{\frac{1}{2}}(z_3-\zeta)}\\
&\hspace{2cm}\times\left\langle V(\ket{0}_{R,\text{rest}},z_1)V(\psi^3_0\ket{0}_{R,\text{rest}},z_2)\right\rangle\Braket{X}_{\text{bos}}\,.
\label{eq:worldsheet result for w=(2,1,1) R sector}
\end{aligned}
\end{equation}
Again, $\braket{X}_{\text{bos}}$ denotes the bosonic contribution of $\braket{X}$. Hence, the worldsheet amplitude again has the form of a correlator in a deformed symmetric product orbifold theory.
Note also that the factor
\begin{equation}
\frac{1}{(z_1-\zeta)^{\frac{3}{2}}(z_2-\zeta)^{\frac{1}{2}}(z_3-\zeta)}\,,
\end{equation}
is expected from the fact that the perturbing field is marginal.

Let us make one observation. Notice that the fermion correlator in this case is precisely the same as \eqref{eq:degenerate R sector fermion} \emph{before} substituting in the covering map data. The fact that the two expressions are the same is consistent with the fact that the vertex operators in those correlators are precisely the same in the fermion theory, see \eqref{eq:spectral flow in R sector with no x translation}. The difference that distinguishes the two in the complete worldsheet amplitude lies in the $\mathfrak{sl}(2,\RR)_{k+2}$ factor where different spectral flow leads to different behaviour of $\gamma$ and thus $x_i$ depend differently on the covering map data.

\section{\boldmath To the bulk of \texorpdfstring{$\rm AdS_3$}{AdS3} in the tensionless limit}\label{sec:tensionless limit}
So far in our discussion we have been in a strict near-boundary limit where we drop the interaction term $\beta\bar\beta e^{-Q\Phi}$. Taking into account this term at generic level $k$ is technically complicated and we currently do not know how to proceed. However, there is a special value of $k$ and a compact CFT $X$ that allows us to investigate the effects of the interaction term in details. This is the tensionless limit $k=1$ string theory in $\rm AdS_3\times S^3\times\mathbb{T}^4$. We will see in this section that perturbatively including the interaction term in the tensionless worldsheet amplitude does not give rise to any corrections to the near-boundary results. In other words, the localisation exhibited in the near-boundary approximation becomes perturbatively exact in the tensionless limit. This essentially gives a perturbative proof to the speculations in \cite{McStay:2023thk,Dei:2023ivl}.\footnote{We should note that the speculations that the interaction does not affect worldsheet correlation functions in \cite{McStay:2023thk,Dei:2023ivl} were formulated in the hybrid formulation of superstring theory as compared to the RNS treatment we adopt here. Assuming that the two prescriptions are equivalent, the same result should hold in the hybrid formulation but maybe with a slightly different proof.}

The full action, including the interaction term, that describes bosonic $\mathfrak{sl}(2,\RR)_{k+2}$ WZW model is
\begin{equation}
\begin{aligned}
S_{\rm AdS_3}=\frac{1}{2\pi}\int d^2z\left( \frac{1}{2}\partial\Phi\bar\partial\Phi-\frac{QR\Phi}{4}+\beta\bar\partial\gamma+\bar\beta\partial\bar\gamma-\nu\beta\bar\beta e^{-Q\Phi}\right)-p\mathcal{O}\,.
\end{aligned}
\end{equation}
Perturbatively expanding in powers of $\nu,p$ and focusing on the term of order $\nu^Mp^{N_M}$, we obtain
\begin{equation}
\begin{aligned}
\left.\Braket{\prod_{i=1}^nV^{w_i}_{m_i,j_i}(z_i,x_i)}_{p,\nu}\right|_{\nu^Mp^{N_M}}\simeq&\int\prod_{c=1}^Md\mu_c\prod_{a=1}^{N_M}d\lambda_a\\
&\Braket{\prod_{c=1}^M\left(\beta e^{-Q\Phi}\right)(\mu_c)\prod_{a=1}^{N_M}D(\lambda_a)\prod_{i=1}^nV^{w_i}_{m_i,j_i}(z_i,x_i)}\,.
\label{eq:order nu^Mp^(N_M) ws correlator}
\end{aligned}
\end{equation}
The integrand of \eqref{eq:order nu^Mp^(N_M) ws correlator} is computed in the Appendix \ref{sec:betagamma with interaction} and the result is \eqref{eq:betagamma correlator with interaction term}
\begin{equation}
\begin{aligned}
&\int\prod_{a=1}^{N_M}dc_a\prod_{c=1}^Mda_cdb\prod_{c=1}^M\delta'(a_c) \prod_{a=1}^{N_M}\left(c_a\right)^{1-k_b}\prod_{i=1}^{n}\left(\frac{\partial^{w_i}(\gamma(z_i)-x_i)}{w_i!}\right)^{-m_i-j_i}\delta_{w_i}(\gamma(z_i)-x_i)\Braket{\Phi}\,,
\label{}
\end{aligned}
\end{equation}
where $c_a,a_c$ are residues of $\gamma$ at $\lambda_a,\mu_c$ respectively and $\Braket{\Phi}$ denotes
\begin{equation}
\Braket{\Phi}=\Braket{\prod_{c=1}^Me^{-Q\Phi}(\mu_c)\prod_{a=1}^{N_M}e^{-2\Phi/Q}(\lambda_a)\prod_{i=1}^{n}e^{-(Qj_i-w_i/Q)\Phi}(z_i)}\,.
\end{equation}
We will, however, not need an explicit form of $\Braket{\Phi}$ in order to complete our proof and the only important piece from $\braket{\Phi}$ is its momentum conservation delta function. From the conservation law \eqref{eq:conservation of Phi}, we have
\begin{equation}
N_M+M=1+\sum_{i=1}^{n}\frac{w_i-1}{2}-\left(\sum_{i=1}^{n}j_i-\frac{n}{2}\right)\,,
\label{eq:Riemann Hurwitz with explicit j}
\end{equation}
where we have specialised to $k=1$ ($k_b=3$) and used $Q=\sqrt{2/k}=\sqrt{2}$. However, it was argued in \cite{Dei:2020zui,Eberhardt:2018ouy,Dei:2023ivl,Eberhardt:2019ywk} that only the continuous representation with $j=\tfrac{1}{2}$ survives in the tensionless limit of strings in $\rm AdS_3\times S^3\times \mathbb{T}^4$. We therefore see that the term in the last bracket actually vanishes. Hence
\begin{equation}
N_M+M=:N=1+\sum_{i=1}^{n}\frac{w_i-1}{2}\,,
\label{eq:Riemann Hurwitz}
\end{equation}
where $N$ is the total number of poles in $\gamma$. Note that this is the Riemann-Hurwitz formula. From this formula, we note that we have
\begin{equation}
2N+1+(n-3)=2(N_M+M)+1+(n-3)=\sum_iw_i\,.
\label{eq:integration equals delta functions}
\end{equation}
However, notice that $2(N_M+M)+1+(n-3)$ is precisely the total number of integrations in the worldsheet amplitude since we have to integrate over $\lambda_a,c_a,\mu_c,a_c,b$ and over the worldsheet moduli space $\mathcal{M}_{0,n}$. Hence, \eqref{eq:integration equals delta functions} implies that we have the same number of integrals as the number of delta functions $\delta_{w_i}(\gamma(z_i)-x_i)$ and this allows us to perform an integration
\begin{equation}
\int d\mathcal{M}_{0,n}\prod_{a=1}^{N_M}d\lambda_adc_a\prod_{c=1}^Md\mu_cda_cdb\prod_{i=1}^n\delta_{w_i}(\gamma(z_i)-x_i)=\mathcal{J}\,,
\label{eq:Jacobian}
\end{equation}
where $\mathcal{J}$ is some non-vanishing expression and exists whenever a covering map $\gamma$ exists. The Jacobian $\mathcal{J}$ has been computed explicitly in \cite{Knighton:2024qxd} but we shall not need its expression. The worldsheet amplitude now becomes
\begin{equation}
\mathcal{J}\prod_{c=1}^M\delta'(a_c) \prod_{a=1}^{N_M}\left(c_a\right)^{1-k_b}\prod_{i=1}^{n}\left(\frac{\partial^{w_i}(\gamma(z_i)-x_i)}{w_i!}\right)^{-m_i-j_i}\Braket{\Phi}\,.
\label{ws amplitude with interaction at genus 0}
\end{equation}
Note that the amplitude contains derivatives of delta function. Furthermore, these derivatives impose the condition that the residues at $\mu_c$ are actually zero, that is, $a_c=0$. However, for a covering map to exist it needs a certain number of poles given by \eqref{eq:Riemann Hurwitz} and imposing the conditions from derivatives of delta function means that the (would-be) covering maps have less poles than needed by \eqref{eq:Riemann Hurwitz} and hence, cannot exist in general. This then implies an important conclusion: the amplitude \eqref{ws amplitude with interaction at genus 0} \emph{generically vanishes}. The only amplitudes that do not vanish in general are the ones that have no interaction term inserted, i.e., those with $M=0$ since they do not have extra conditions coming from the $\delta'(a_c)$.

To show that an analogous statement holds even at higher genus, we first note that \eqref{eq:Riemann Hurwitz} changes to
\begin{equation}
N_M+M=1-g+\sum_{i=1}^{n}\frac{w_i-1}{2}-\left(\sum_{i=1}^{n}j_i-\frac{n}{2}\right)\,,
\label{eq:Riemann Hurwitz at genus g}
\end{equation}
and with the restriction $j=\tfrac{1}{2}$ on the allowed spectrum, we again have
\begin{equation}
N_M+M=N=1-g+\sum_{i=1}^{n}\frac{w_i-1}{2}\,.
\end{equation}
This leads to
\begin{equation}
(2N+1-g)+(n+3g-3)=\sum_iw_i\,.
\label{eq:integration equals delta functions at genus g}
\end{equation}
From the analysis of \cite{Knighton:2023mhq}, the worldsheet amplitude consists of integrals over $\mathcal{M}_{g,n}$ and integrals over a space of meromorphic functions with $N$ simple poles and the dimensions of these integrals are $n+3g-3$ and $2N+1-g$, respectively. Thus \eqref{eq:integration equals delta functions at genus g} tells us again that the number of integrations is the same as the number of delta functions $\delta_{w_i}(\gamma(z_i)-x_i)$ and the worldsheet amplitude takes the form \eqref{ws amplitude with interaction at genus 0} with $\mathcal{J}$ replaced by its higher genus analogue. The derivatives of delta functions again impose that the number poles are less than what is needed for a covering map to exist in general and we conclude that the worldsheet amplitude with interaction term inserted also vanishes generically at higher genus.

Therefore, we conclude that inserting any positive number of the interaction terms leads to vanishing worldsheet amplitudes and this is valid for $n$-point functions at arbitrary genus. This implies that the interaction term does not contribute perturbatively in the tensionless limit. We further conclude that using the free action \eqref{eq:action} or the full action \eqref{eq:ads full action} are equivalent perturbatively and thus, the near-boundary approximation becomes perturbatively exact in the tensionless limit. However, we should point out that the non-perturbative effects are not captured by our perturbative analysis. Nevertheless, we expect the non-perturbative effects to be trivial and the near-boundary limit to be exact beyond perturbation theory.

\section{Discussion and Future Directions}\label{sec:discussion}

\subsection{Discussion and summary}

Let us start by summarising our results. Firstly, we have computed a number of 3- and 4-point superstring amplitudes using the near-boundary approximation and we showed that they can be written as correlators in a certain deformed symmetric orbifold. Our results imply that for superstrings propagating in the background ${\rm AdS_3}\times X$, the dual CFT is a deformed symmetric orbifold 
\begin{equation}
{\rm Sym}^\infty\left(\RR^{(1)}_\mathcal{Q}\times X\right)+\mu\int \sigma_{2,\alpha,\Psi_X}\,,
\label{eq:conjectured dual CFT}
\end{equation}
where $\sigma_{2,\alpha,\Psi_X}$ denotes a (possibly supercurrent descendant) marginal perturbation in the $2$-cycle twisted sector which carries dilaton momentum $\alpha$, see equation \eqref{eq:linear dilaton momentum}. $\Psi_X$ denotes the remaining piece of the perturbation that depends on $X$. Although we have not determined the form of $\Psi_X$ explicitly, we believe that given our demonstration in this paper it should be within reach once $X$ is specified. Finally, we have discussed the effect of perturbatively including the interaction term $\beta\bar\beta e^{-Q\Phi}$ in the tensionless limit. We showed that $n$-point, genus $g$ amplitudes of spectrally flowed ground states receive no perturbative corrections from the interaction term. We thus essentially prove the claim speculated in \cite{McStay:2023thk,Dei:2023ivl} that this term can be ignored in the tensionless limit, at least at the perturbative level.

We now discuss how our results are related to various other results in the literature. We start by noting that our proposed dual CFT is consistent with the $k=1$ proposal in \cite{Giribet:2018ada} which is the free symmetric orbifold
\begin{equation}
{\rm Sym}^N\left(\RR^{(1)}\times \tilde X\right)\,,
\end{equation}
where $\tilde X={\rm S^1 \times(\text{ 6 free fermions}})$ where we emphasise that the $\rm S^1$ factor is an $\mathcal{N}=1$ supersymmetric theory. Let us see in more details how this agreement comes about. Starting from our proposal \eqref{eq:conjectured dual CFT} and focusing on the continuous representations (long strings),\footnote{In \cite{Giribet:2018ada}, the authors note that the long string spectrum is sufficient to capture the symmetric orbifold spectrum.} we deduce that the sum of $SL(2,\RR)$ spins is of the form
\begin{equation}
\sum_{i=1}^nj_i=\frac{n}{2}+i\sum_{i=1}^ns_i\,,
\label{eq:reality of sum of j_i}
\end{equation}
where $s_i\in\RR$ for all $i$'s. However, we see from \eqref{eq:generalised j constraint} with $k_b=k+2=3$ that $\sum_is_i=m=0$ since $\sum_ij_i$ is constrained to be real. Hence, we conclude that all non-vanishing near-boundary amplitudes satisfy the $j$-constraint \eqref{eq:j constraint}. From the discussions in \cite{Knighton:2024qxd} and Section \ref{sec:bosonic correlation functions}, we then conclude that near-boundary worldsheet amplitudes reproduce correlators in the free symmetric orbifold ${\rm Sym}^N\left(\RR^{(1)}_\mathcal{Q}\times \tilde X\right)$ and that this is the dual CFT in this case. Our proposed CFT \eqref{eq:conjectured dual CFT} also narrows down the candidate for the seed theory in \cite{Argurio:2000tb} which was proposed to be some $c=6k$ theory. Here, we see that the seed theory should be $\RR^{(1)}_\mathcal{Q}\times X$ for superstring theory in ${\rm AdS_3}\times X$.

If we now apply our result to the case $X={\rm S^3}\times\mathcal{M}_4$ with $\mathcal{M}_4=K3$ or $\mathbb{T}^4$, we recover the conjecture of \cite{Eberhardt:2021vsx} as follows. The CFT describing $X$ is a tensor product $\mathfrak{su}(2)^{(1)}_{k}\otimes\mathcal{M}_4$, where the level of the $\mathfrak{su}(2)^{(1)}$ theory is the same as the level in the $\mathfrak{sl}(2,\RR)^{(1)}$ theory. Thus, from our correlator computations previously, a tentative dual CFT is\footnote{We suspect that the theory $\mathbb{R}_{\mathcal{Q}}^{(1)}\times \mathfrak{su}(2)^{(1)}_{k}$ posses an $\mathcal{N}=4$ supersymmetry but we currently cannot write down explicitly $\mathcal{N}=4$ generators.}
\begin{equation}\label{eq:string-duality}
\begin{gathered}
\text{Sym}^\infty\left(\mathbb{R}_{\mathcal{Q}}^{(1)}\times \mathfrak{su}(2)^{(1)}_{k}\times\mathcal{M}_4\right)+\mu\int\sigma_{2,\alpha,\Psi_{\mathfrak{su}(2)^{(1)}_{k}\times\mathcal{M}_4}}\,.
\end{gathered}
\end{equation}
This duality \eqref{eq:string-duality} is the proposal of \cite{Eberhardt:2021vsx}. Furthermore, it was noted in \cite{Eberhardt:2021vsx} that the seed theory has the right spectrum of delta-function normalisable operators and hence, should be the correct seed theory. Note also that in the tensionless limit $k=1$, the central charge of the theory $\mathbb{R}_{\mathcal{Q}}^{(1)}\times \mathfrak{su}(2)^{(1)}_{k}$ is zero. It was then argued in \cite{Eberhardt:2021vsx} that the theory $\mathbb{R}_{\mathcal{Q}}^{(1)}\times \mathfrak{su}(2)^{(1)}_{k}$ is trivial at $k=1$. Note further that since only the states with $j=\tfrac{1}{2}$ survives at $k=1$, the $j$-constraint \eqref{eq:j constraint} is never violated and the dual CFT is therefore free. In summary, the dual CFT at $k=1$ is 
\begin{equation}\label{eq:string-duality at k=1}
\begin{gathered}
\text{Sym}^\infty\left(\mathcal{M}_4\right)\,.
\end{gathered}
\end{equation}
Therefore, our results in this paper give another piece of evidence supporting the duality proposed in \cite{Eberhardt:2021vsx}.

There is another interesting $\mathcal{M}_4$ manifold which is $\rm S^3\times S^1$. In this case, the string physical spectrum was studied in \cite{Eberhardt:2019niq,Eberhardt:2017pty,Eberhardt:2017fsi}. It was argued in \cite{Eberhardt:2019niq} that the dual CFT in this case is a symmetric orbifold of a large $\mathcal{N}=4$ Liouville theory. However, we see from our computations that a more natural candidate is
\begin{equation}\label{eq:string-duality for S^3xS^1}
\begin{gathered}
\text{Sym}^\infty\left(\mathbb{R}_{\mathcal{Q}}^{(1)}\times \mathfrak{su}(2)^{(1)}_{k^+}\times\mathfrak{su}(2)^{(1)}_{k^-}\times\mathfrak{u}(1)^{(1)}\right)+\mu\int\sigma_{2,\alpha,{\rm S^3\times S^3\times S^1}}\,.
\end{gathered}
\end{equation}
We propose that \eqref{eq:string-duality for S^3xS^1} is the dual CFT for $k^\pm\geq2$. What's more, the seed theory has the correct central charge which is 
\begin{equation}
6k=\frac{6k^+k^-}{k^++k^-}
\end{equation}
where we have used
\begin{equation}
\frac{1}{k}=\frac{1}{k^+}+\frac{1}{k^-}\,,
\end{equation}
which follows from superstring criticality. 
The situation when the flux through one of the spheres is minimal, say $k^+=1$, can also be analysed but is more speculative than our analysis so far. 
Extrapolating our result \eqref{eq:string-duality for S^3xS^1} to $k^+=1$, one may propose that the dual CFT is
\begin{equation}\label{eq:string-duality for S^3xS^1 and k^+=1}
\begin{gathered}
\text{Sym}^\infty\left(\mathbb{R}_{\mathcal{Q}}^{(1)}\times \mathfrak{su}(2)^{(1)}_{1}\times\mathfrak{su}(2)^{(1)}_{k^-}\times\mathfrak{u}(1)^{(1)}\right)+\mu\int\sigma_{2,\alpha,{\rm S^3\times S^3\times S^1}}\,.
\end{gathered}
\end{equation}
However, the seed theory in this case contains a non-unitary factor $\mathfrak{su}(2)^{(1)}_1$. Following the discussion in \cite{Eberhardt:2021vsx}, we suspect that the combination $\mathbb{R}_{\mathcal{Q}}^{(1)}\times \mathfrak{su}(2)^{(1)}_{1}\times\mathfrak{su}(2)^{(1)}_{k^-}\times\mathfrak{u}(1)^{(1)}$ has a unitary limit as $k^+\to1$, this will be discussed in detail elsewhere \cite{Gaberdiel:2024abc}. Lastly, we would like to note that our general proposal \eqref{eq:conjectured dual CFT} is also consistent with the squashed sphere proposal and other proposal discussed around eq.(1.3-1.4) in \cite{Balthazar:2021xeh}.

\subsection{Future directions}
We end our exposition with various interesting future directions.

\paragraph{\boldmath Superstrings in \texorpdfstring{$\rm AdS_3\times S^3\times\mathcal{M}_4$}{AdS3 times S3 times M4}:}
We have seen that it is possible to calculate correlators in the near-boundary limit of $\rm AdS_3$. In this paper, we make our discussion general and therefore we have not considered a specific background or compactification in detail. It would be interesting to perform correlator computations in more details in a specific background. In particular for the background $\mathcal{M}_4=\mathbb{T}^4,K3$ and ${\rm S^3\times S^1}$, it would be useful to explicitly determine the perturbation on the CFT side and for the case $\mathcal{M}_4=\mathbb{T}^4,K3$ compare with the proposed perturbation of \cite{Eberhardt:2021vsx}. It would also be interesting to explore tensionless limit in the RNS language in more detail since most progress comes from using the hybrid formulation, see \cite{Dei:2023ivl,Gaberdiel:2022als,Dei:2020zui,Gaberdiel:2021njm,Fiset:2022erp,Gaberdiel:2022oeu,Gaberdiel:2021kkp} and references therein.

\paragraph{Turning on infinitesimal RR flux:}
The next thing that should be within reach is the perturbative study of RR flux. 
It would be interesting to understand how to turn on the RR flux and study the effects turning on the RR flux has on the duality \eqref{eq:conjectured dual CFT}. In particular in the tensionless limit, when the dual CFT is free at the pure NSNS point, it was identified in \cite{Fiset:2022erp} a dual perturbation needed to turn on when the RR flux is switched on on the worldsheet. However, the computations in \cite{Fiset:2022erp} were done using the hybrid formalism and it will be very useful to understand this perturbation in a more conventional RNS formalism and also to understand to the RR flux at higher $k$.

\paragraph{\boldmath Non-perturbative study of the \texorpdfstring{$\rm AdS_3$}{AdS3} interaction term:}
We have shown that the interaction term $\beta\bar\beta e^{-Q\Phi}$ does not contribute perturbatively in the tensionless limit. However, to completely prove the claim in \cite{McStay:2023thk,Dei:2023ivl}, non-perturbative treatment of the interaction term is needed to ensure that the free field approximation is valid even in the non-perturbative regime.

\paragraph{Higher genus correlators:}
Given the success of the computations at genus zero both in the bosonic string and superstring theories, it is natural to consider higher genus correlators next. The most nontrivial calculation is the calculation of Wakimoto $\beta\gamma$ correlators. Given this correlator for a generic $n$-point function, the remaining parts of a correlator are simply calculated by Wick contractions in higher genus. It would then be fruitful to see if it is possible to derive $1/N$ corrections to CFT correlators directly from the worldsheet of higher genus. 

\acknowledgments
I thank Cassiano Daniel, Matthias Gaberdiel, Bob Knighton, Nathan McStay, Kiarash Naderi and Beat Nairz for useful discussions. I also thank Matthias Gaberdiel, Bob Knighton and Beat Nairz for comments on a draft version of the paper. I would like to specially thank Matthias Gaberdiel and Bob Knighton for invaluable and fruitful discussions that inspired this work and at various stages of this work. The work of VS is supported by a grant from the Swiss National Science Foundation. The activities of the group are more generally supported by the NCCR SwissMAP, which is also funded by the Swiss National Science Foundation. 

\appendix

\section{Assorted calculations}\label{sec:all tedious calculations}

\subsection{Fermion correlators in Case III}\label{sec:tedious calculation in case III}
The fermion correlator in \eqref{eq:1st contribution} can be calculated as follows. Wick contractions give
\begin{equation}
\begin{aligned}
\frac{x_{12}^2x_{34}^4k^3}{z_{12}z_{34}^4}-\frac{z_{23}x_{24}^2x_{34}^2x_{13}^2k^3}{z_{13}^2z_{24}^2z_{34}^2}+\frac{x_{14}^2x_{34}^2x_{23}^2z_{13}k^3}{z_{23}^2z_{34}^2z_{14}^2}+\frac{k^3x_{13}^2x_{34}^2x_{24}^2}{z_{13}}\left( -\frac{2}{z_{24}z_{34}^3}+\frac{1}{z_{24}^2z_{34}^2} \right)&\\
-\frac{k^3x_{34}^2x_{14}^2x_{23}^2}{z_{23}}\left(- \frac{2}{z_{34}^3z_{14}}+\frac{1}{z_{34}^2z_{14}^2} \right)&\,.
\end{aligned}
\end{equation}
Using an analogue of \eqref{eq:covering map data w=(1,1,1)} in the case where $w=(1,1,1,1)$, we have
\begin{equation}
\begin{aligned}
x_{ij}=-\frac{c z_{ij}}{(z_i-\lambda)(z_j-\lambda)},\quad C^\gamma=-c,\quad a_i=-\frac{c}{(z_i-\lambda)^2}\,.
\label{eq:covering map data w=(1,1,1,1)}
\end{aligned}
\end{equation}
Thus, we can rewrite the fermion contribution as
\begin{equation}
\begin{aligned}
&\frac{k^3(C^\gamma)^6}{(z_1-\lambda)^2(z_2-\lambda)^2(z_3-\lambda)^4(z_4-\lambda)^4}\times\left( \frac{z_{12}^2z_{34}^4}{z_{12}z_{34}^4}-\frac{z_{23}z_{24}^2z_{34}^2z_{13}^2}{z_{13}^2z_{24}^2z_{34}^2}+\frac{z_{14}^2z_{34}^2z_{23}^2z_{13}}{z_{14}^2z_{23}^2z_{34}^2}\right.\\
&\hspace{3cm}\left.+\frac{z_{13}^2z_{34}^2z_{24}^2}{z_{13}}\left( 
\frac{1}{z_{24}^2z_{34}^2}-\frac{2}{z_{24}z_{34}^3} \right)-\frac{z_{34}^2z_{14}^2z_{23}^2}{z_{23}}\left( \frac{1}{z_{14}^2z_{34}^2}-\frac{2}{z_{34}^3z_{14}} \right) \right)\\
&=k^3a_1a_2a_3^2a_4^2z_{12}\,.
\end{aligned}
\end{equation}
The fermion contribution of \eqref{eq:2nd and 3rd contribution} is given by
\begin{equation}
\begin{aligned}
&\langle\delta_1(\psi^-,x_1)\delta_1(\psi^-,x_2)(\psi^3-x_3\psi^+)\delta_1(\psi^-,x_3)\delta_2(\psi^-,x_4)\rangle\\
&=\frac{k^3x_{14}^2x_{23}x_{34}x_{24}z_{13}}{z_{23}z_{14}^2z_{34}^2}-\frac{k^3x_{24}^2x_{13}x_{34}x_{14}z_{23}}{z_{13}z_{24}^2z_{34}^2}\,.
\end{aligned}
\end{equation}
Using \eqref{eq:covering map data w=(1,1,1,1)}, we obtain
\begin{equation}
\begin{aligned}
\frac{k^3x_{14}^2x_{23}x_{34}x_{24}z_{13}}{z_{23}z_{14}^2z_{34}^2}-\frac{k^3x_{24}^2x_{13}x_{34}x_{14}z_{23}}{z_{13}z_{24}^2z_{34}^2}=k^3a_1a_2a_3a_4^2z_{12}\,.
\end{aligned}
\end{equation}
The fermions in \eqref{eq:4th contribution} contributes
\begin{equation}
\begin{aligned}
&\langle\delta_1(\psi^-,x_1)\delta_1(\psi^-,x_2)(\psi^3-x_3\psi^+)\delta_1(\psi^-,x_3)(\psi^3-x_4\psi^+)\delta_1(\psi^-,x_4)\rangle\\
&=-\frac{k^3x_{34}^2x_{12}^2}{z_{12}z_{34}^2}+\frac{k^3x_{13}x_{24}}{2z_{13}z_{24}z_{34}}(2x_{12}x_{34}-x_{13}x_{24})+\frac{k^3x_{14}x_{23}}{z_{14}z_{23}z_{34}}(2x_{12}x_{34}+x_{14}x_{23})\,.
\end{aligned}
\end{equation}
Using \eqref{eq:covering map data w=(1,1,1,1)}, we obtain
\begin{equation}
\begin{aligned}
&\langle\delta_1(\psi^-,x_1)\delta_1(\psi^-,x_2)(\psi^3-x_3\psi^+)\delta_1(\psi^-,x_3)(\psi^3-x_4\psi^+)\delta_1(\psi^-,x_4)\rangle\\
&=k^3a_1a_2a_3a_4z_{12}\,.
\end{aligned}
\end{equation}
To show that \eqref{eq:zero correlator} is actually zero, we note that the fermion correlator in \eqref{eq:zero correlator} gives
\begin{equation}
\begin{aligned}
&\Braket{(\psi^--2x_1\psi^3+x_1^2\psi^+)(\psi^--2x_2\psi^3+x_2^2\psi^+)(\psi^--2x_3\psi^3+x_3^2\psi^+)(\psi^--2x_4\psi^3+x_4^2\psi^+)}\\
&=\frac{x_{12}^2x_{34}^2}{z_{12}z_{34}}-\frac{x_{13}^2x_{24}^2}{z_{13}z_{24}}+\frac{x_{14}^2x_{23}^2}{z_{14}z_{23}}\,.
\end{aligned}
\end{equation}
Using \eqref{eq:covering map data w=(1,1,1,1)}, we obtain
\begin{equation}
\begin{aligned}
\frac{x_{12}^2x_{34}^2}{z_{12}z_{34}}-\frac{x_{13}^2x_{24}^2}{z_{13}z_{24}}+\frac{x_{14}^2x_{23}^2}{z_{14}z_{23}}=&\frac{c^4}{(z_1-\lambda)^2(z_2-\lambda)^2(z_3-\lambda)^2(z_4-\lambda)^2}\left( z_{12}z_{34}-z_{13}z_{24}+z_{14}z_{23} \right)\\
=&0\,.
\end{aligned}
\end{equation}

\subsection{Fermion correlators in Case IV}\label{sec:tedious calculation in case IV}
The fermion contribution from the first term of \eqref{eq:full sl(2,R) w=(2,2,1)} is
\begin{equation}
\begin{aligned}
&\langle\delta_2(\psi^-,x_1)\delta_2(\psi^-,x_2)\delta_2(\psi^-,x_3)\rangle=\frac{4k^3x_{12}^2x_{13}^2x_{23}^2}{z_{12}^2z_{13}^2z_{23}^2}\,,
\end{aligned}
\end{equation}
while the fermion contribution from the second term in \eqref{eq:full sl(2,R) w=(2,2,1)} is
\begin{equation}
\begin{aligned}
\langle\delta_2(\psi^-,x_1)\delta_2(\psi^-,x_2)\rangle=\frac{k^2x_{12}^4}{z_{12}^4}\,.
\end{aligned}
\end{equation}
Lastly, the fermion correlator in the last term in \eqref{eq:full sl(2,R) w=(2,2,1)} contributes
\begin{equation}
\begin{aligned}
&\langle\delta_2(\psi^-,x_1)\delta_2(\psi^-,x_2)(\psi^3-x_3\psi^+)\delta_1(\psi^-,x_3)\rangle=\frac{2k^3x_{12}^3x_{13}x_{23}}{z_{12}^3z_{13}z_{23}}\,.
\end{aligned}
\end{equation}
From \eqref{eq:degree with j constraint}, we have $N=2$ in this case. Hence, the $\gamma$ field is of the form
\begin{equation}
\gamma(z)=b+\frac{c_1}{z-\lambda_1}+\frac{c_2}{z-\lambda_2}\,,
\end{equation}
with $b$ constant. We therefore have from \eqref{eq:definition of a_i} and \eqref{eq:definition of C^gamma} that
\begin{equation}
\begin{aligned}
c_1=&-\frac{C^\gamma(z_1-\lambda_1)(z_2-\lambda_1)}{\lambda_{12}^2},&\quad c_2=&-\frac{C^\gamma(z_1-\lambda_2)(z_2-\lambda_2)}{\lambda_{12}^2}\,,\\
2a_1=&\frac{C^\gamma z_{12}}{(z_1-\lambda_1)^2(z_1-\lambda_2)^2},&\quad 2a_2=&-\frac{C^\gamma z_{12}}{(z_2-\lambda_1)^2(z_2-\lambda_2)^2}\,,\\
a_3=&\frac{C^\gamma z_{13}z_{23}}{(z_3-\lambda_1)^2(z_3-\lambda_2)^2},&\quad x_{12}=&-\frac{C^\gamma z_{12}^2}{\lambda_{12}(z_1-\lambda_2)(z_2-\lambda_1)}\,,\\
x_{13}=&-\frac{C^\gamma z_{13}^2(z_2-\lambda_2)}{\lambda_{12}(z_1-\lambda_2)(z_3-\lambda_1)(z_3-\lambda_2)},&\quad x_{23}=&-\frac{C^\gamma z_{23}^2(z_1-\lambda_2)}{\lambda_{12}(z_2-\lambda_2)(z_3-\lambda_1)(z_3-\lambda_2)}\,,\\
1=&-\frac{(z_1-\lambda_2)(z_2-\lambda_1)}{(z_1-\lambda_1)(z_2-\lambda_2)}\,.
\end{aligned}
\end{equation}
Hence, the fermion contributions can be rewritten as
\begin{equation}
\begin{aligned}
\frac{4x_{12}^2x_{13}^2x_{23}^2k^3}{z_{12}^2z_{13}^2z_{23}^2}\simeq&\frac{c_1c_2z_{12}(a_1a_2)^{\frac{3}{2}}(a_3)^2}{C^\gamma}\,,\\
\frac{k^2x_{12}^4}{z_{12}^4}\simeq&\frac{c_1c_2z_{12}(a_1a_2)^{\frac{3}{2}}}{C^\gamma}\,,\\
\frac{2k^3x_{12}^3x_{13}x_{23}}{z_{12}^3z_{13}z_{23}}\simeq&\frac{c_1c_2z_{12}(a_1a_2)^{\frac{3}{2}}a_3}{C^\gamma}\,.
\end{aligned}
\end{equation}

\section{\boldmath \texorpdfstring{$\beta\gamma$}{betagamma} path integral with interaction insertions}\label{sec:betagamma with interaction}
In the main text, we have to evaluate the following correlator
\begin{equation}
\Braket{\prod_{c=1}^M\left(\beta e^{-Q\Phi}\right)(\mu_c)\prod_{a=1}^{N_M}D(\lambda_a)\prod_{i=1}^nV^{w_i}_{m_i,j_i}(z_i,x_i)}\,,
\label{eq:wanted correlator}
\end{equation}
which we will now do in this appendix. The $\Phi$ correlator is
\begin{equation}
\begin{aligned}
\Braket{\prod_{c=1}^Me^{-Q\Phi}(\mu_c)\prod_{a=1}^{N_M}e^{-2\Phi/Q}(\lambda_a)\prod_{i=1}^{n}e^{-(Qj_i-w_i/Q)\Phi}(z_i)}\,.
\end{aligned}
\end{equation}
Using Wick contractions, we obtain
\begin{equation}
\begin{aligned}
\prod_{c<d}(\mu_c-\mu_d)^{-Q^2}\prod_{c,a}(\mu_c-\lambda_a)^{-2}\prod_{c,i}(\mu_c-z_i)^{w_i-Q^2j_i}\prod_{a<b}(\lambda_a-\lambda_b)^{-\frac{4}{Q^2}}\prod_{i,a}(z_i-\lambda_a)^{2(w_i/Q^2-j_i)}&\\
\times\prod_{i<j}(z_i-z_j)^{-(w_i/Q-j_iQ)(w_j/Q-j_jQ)}\delta\bigg(MQ+\frac{2N_M}{Q}+\sum_{i=1}^{n}\left(Qj_i-\frac{w_i}{Q}\right)-Q\bigg)&\,.
\label{eq:Phi correlator with interaction term}
\end{aligned}
\end{equation}
The argument of the momentum conservation delta function can be written as
\begin{equation}
N_M=1+\sum_{i=1}^{n}\frac{w_i-1}{2}-\frac{Q^2}{2}\left(\sum_{i=1}^{n}j_i-\frac{k_b}{2}(n-2)+(n-3)\right)-\frac{MQ^2}{2}\,.
\label{eq:conservation of Phi}
\end{equation}
Let us now turn to the $\beta\gamma$ correlator which takes the form
\begin{equation}
\begin{aligned}
\Braket{\prod_{c=1}^M\beta(\mu_c)\prod_{a=1}^{N_M}\left(\left(\oint\gamma\right)^{1-k_b}\delta(\beta) \right)(\lambda_a)\prod_{i=1}^{n}\left( 
\left(\frac{\partial^{w_i}(\gamma(z_i)-x_i)}{w_i!}\right)^{-m_i-j_i}\delta_{w_i}(\gamma(z_i)-x_i) \right)}\,.
\end{aligned}
\end{equation}
Following the discussion in \cite{Knighton:2024qxd}, we write $\delta(\beta(\lambda_a))$ and $\beta(\mu_c)$ as Fourier representations
\begin{equation}
\delta(\beta(\lambda_a))=\int\frac{\mathrm{d}\xi_a}{2\pi}e^{i\beta(\lambda_a)\xi_a}\,,
\end{equation}
as well as
\begin{equation}
\begin{aligned}
\beta(\mu_c)=&i\int d\varsigma_c e^{i\varsigma_c\beta(\mu_c)}\delta'(\varsigma_c)\,.
\end{aligned}
\end{equation}
Hence, we can rewrite our $\beta\gamma$ path integral as
\begin{equation}
\begin{aligned}
&\int\mathcal{D}\beta\mathcal{D}\gamma\prod_{a=1}^{N_M}d\xi_a\prod_{c=1}^Md\varsigma_c\,\\
&e^{-S'[\beta,\gamma]}\prod_{c=1}^M\delta'(\varsigma_c)\prod_{a=1}^{N_M}\left(\oint_{\lambda_a}\gamma\right)^{-(k_b-1)}\prod_{i=1}^{n}\left(\frac{\partial^{w_i}(\gamma(z_i)-x_i)}{w_i!}\right)^{-m_i-j_i}\delta_{w_i}(\gamma(z_i)-x_i)\,,
\end{aligned}
\end{equation}
where
\begin{equation}
S'[\beta,\gamma]:=\frac{1}{2\pi}\int_{\Sigma}d^2z\beta\left(\overline{\partial}\gamma-\sum_a2\pi i \xi_a\delta^{(2)}(z,\lambda_a)-\sum_c2\pi i \varsigma_c\delta^{(2)}(z,\mu_c)\right)\,.
\end{equation}
Integrating out $\beta$, we obtain that $\gamma$ has poles at $\lambda_a,\mu_c$ with residues $\xi_a,\varsigma_c$, respectively. As noted in \cite{Knighton:2024qxd}, the path integral $\mathcal{D}\gamma$ over such meromorphic functions can be simplified to
\begin{equation}
\begin{aligned}
&\int\prod_{a=1}^{N_M}dc_a\prod_{c=1}^Mda_cdb\prod_{c=1}^M\delta'(a_c) \prod_{a=1}^{N_M}\left(c_a\right)^{-(k_b-1)}\prod_{i=1}^{n}\left(\frac{\partial^{w_i}(\gamma(z_i)-x_i)}{w_i!}\right)^{-m_i-j_i}\delta_{w_i}(\gamma(z_i)-x_i)\,,
\label{eq:betagamma correlator with interaction term}
\end{aligned}
\end{equation}
where we have defined $c_a:=\xi_a,a_c:=\varsigma_c$ and thus $\gamma$ now takes the form
\begin{equation}
\gamma(z)=\sum_{a=1}^{N_M}\frac{c_a}{z-\lambda_a}+\sum_{c=1}^M\frac{a_c}{z-\mu_c}+b\,.
\end{equation}
Hence, the correlator \eqref{eq:wanted correlator} is given by a product of \eqref{eq:Phi correlator with interaction term} and \eqref{eq:betagamma correlator with interaction term}. 

\bibliography{bibliography.bib}
\bibliographystyle{utphys.sty}

\end{document}